\documentclass[aps,prb,twocolumn,superscriptaddress,longbibliography,floatfix]{revtex4-2}

\usepackage{amsmath,amssymb,bm}
\usepackage[mathscr]{eucal}
\usepackage[cal=boondoxo]{mathalfa}

\usepackage{graphicx}
\usepackage{dcolumn}
\usepackage{tabularx}
\usepackage{booktabs}
\usepackage{array}
\usepackage{rotating}

\usepackage{soul}
\usepackage{xcolor}
\definecolor{docnotelinkcolor}{RGB}{0,0,255}
\usepackage{enumitem}
\usepackage{hyperref} %
\usepackage{caption}
\usepackage{ragged2e}
\makeatletter
\long\def\@makecaption#1#2{%
 \vskip\abovecaptionskip
 \begingroup\small
 \setbox\@tempboxa\hbox{\textbf{#1.} #2}%
 \begin{minipage}{\hsize}\justifying\textbf{#1.} #2\end{minipage}%
 \endgroup
 \vskip\belowcaptionskip}
\makeatother

\renewcommand{\vec}[1]{{\mathbf #1}}

\newcommand{\be}{\begin{equation}}
\newcommand{\ee}{\end{equation}}

\definecolor{azure}{rgb}{0.05,0.55,0.85}

\usepackage[normalem]{ulem}
\usepackage{needspace} %

\begin{document}

\title{A unified sensitivity kernel for coherent wave tomography \\
in radar, seismic, ultrasound, and optical imaging}

\author{Alexey Yamilov}
\email{yamilov@mst.edu}
\affiliation{Department of Physics, Missouri University of Science and Technology, Rolla, Missouri 65409, USA}

\date{\today}

\begin{abstract}
Imaging an inhomogeneous medium with waves rests on the response of a boundary measurement to a change of the medium inside, formally the sensitivity kernel of the inverse problem.
Here we show that the kernels of radar, seismology, medical ultrasound, and diffuse optics all reduce to just two classes.
Every kernel is the product of a forward and an adjoint field, and a real measurement retains one quadrature of that product.
These quadratures are the two classes, selected either by a component of the material parameter or by the misfit, the structure that persists for the coherent-wave kernel in both the ballistic and the diffusive regime.
We place these areas on one diagram of scattering strength and transport depth, exposing a shared computational limit set by wavefield storage, and propose a broadband on-shell compression that can relax that limit.
\end{abstract}

\maketitle

\section{Introduction \label{sec:introduction}}

Seismic exploration applies full-waveform inversion (FWI) of elastic waves to resolve subsurface wave-speed structure~\cite{1984_Tarantola_FWI,2009_Virieux_FWI_review,2011_Fichtner_FWI_book}, while tomographic synthetic-aperture radar (TomoSAR) recovers volumetric reflectivity of forest canopies from L- and P-band microwave acquisitions through multi-baseline spectral estimation along elevation~\cite{2000_Reigber_Moreira_TomoSAR,2012_Tebaldini_Rocca_TomoSAR}.
Ground-penetrating radar (GPR) maps the shallow subsurface with lower-frequency electromagnetic pulses~\cite{2007_Ernst_GPR_FWI,2019_Klotzsche_GPR_FWI_review}, whereas
medical ultrasound uses MHz acoustic arrays to image soft tissue and transcranial regions~\cite{2019_Wiskin_breast_ultrasound,2020_Guasch_FWI_brain}.
These areas operate in the quasi-ballistic regime, where the wave nature of propagation is essential for the inversion despite the distinct physics of the underlying elastic, electromagnetic, and acoustic fields.
In contrast to these quasi-ballistic areas, diffuse optical tomography (DOT) uses near-infrared (NIR) light for deep imaging of biological tissues in the diffusive regime, where the inverse problem can be recast without an explicit wave description as a random walk of photons~\cite{1999_Arridge_optical_tomography,2010_Durduran_diffuse_optics}.
Wavefront-shaped DOT (WFS-DOT) has recently been proposed to exploit the wave nature of NIR light even in this diffusive regime via a deterministic, sample-dependent input wavefront~\cite{2025_Jara_coherent_wave_sensing,2012_Mosk_controlling_waves}.
With the wave description restored, WFS-DOT joins the quasi-ballistic four as the fifth area considered in this work.

On a fundamental level, each of these areas recovers tomographic information about the material parameter of a wave equation.
In general, this parameter is complex~\cite{1969_Wolf_diffraction_tomography,1988_Kak_Slaney_book}, its real part governing the wave speed and its imaginary part the loss.
The inverse problem is commonly solved iteratively, via the gradient of a data misfit~\cite{1984_Tarantola_FWI,1999_Arridge_optical_tomography}.
This gradient is the residual-weighted counterpart of the sensitivity kernel, the functional derivative of the measured signal with respect to the local value of the material parameter at position $\vec r_0$~\cite{1967_Backus_Gilbert,1995_Arridge_PMDF,1999_Marquering_banana_doughnut}.
In the Born approximation~\cite{1982_Devaney_Born}, the kernel is the product of a forward field $\phi_\mathrm{S}(\vec r_0)$ sent from the source and an adjoint field $\phi_\mathrm{D}(\vec r_0)$ back-propagated from the detector~\cite{1971_Claerbout_imaging,2000_Fink_time_reversal,2005_Tromp_adjoint_banana_doughnut,2025_Jara_coherent_wave_sensing}, both complex at each frequency.
Therefore, two simulations suffice to obtain the kernel at every interior voxel, 
a construction known as the adjoint-state method in seismic FWI~\cite{2006_Plessix_adjoint_state_review} that appears under other names in the areas where it arose independently.
The GPR and medical ultrasound communities adopted it from seismic FWI~\cite{2007_Ernst_GPR_FWI,2015_Sandhu_ultrasound_FWI}.
The adoption happened case by case, at the level of technique.
Each area kept its own formalisms and nomenclature, and, to our knowledge, a common structure spanning all five areas and both transport regimes has not been discussed.

In practice, each area derives several kernels.
Seismology alone catalogs kernels for cross-correlation traveltime, amplitude, envelope, and instantaneous-phase measurements, among others~\cite{2011_Fichtner_FWI_book}.
The finite-frequency traveltime kernel is referred to as `banana--doughnut' after its hollow shape~\cite{1999_Marquering_banana_doughnut,2000_Dahlen_banana_doughnut}, which has been attributed to its Fresnel-zone sensitivity~\cite{1991_Luo_Schuster_traveltime,2004_Spetzler_Snieder_Fresnel}.
Ref.~\cite{2002_Dahlen_Baig_amplitude} derived the traveltime and amplitude kernels side by side, linking them to the sine and the cosine of the excess-path phase.
GPR jointly recovers permittivity and conductivity~\cite{2014_Lavoue_GPR_FWI}, while ultrasound and seismology recover wave speed and attenuation~\cite{2019_Wiskin_breast_ultrasound,2005_Tromp_adjoint_banana_doughnut}.
In visco-acoustic FWI, wave speed and attenuation are treated as two `parameter classes', and their Fr\'echet derivatives are known to differ by a $90^\circ$ phase shift~\cite{2013_Kamei_Pratt_viscoacoustic_FWI}.

Here we obtain three results.
First, we show that the entire kernel catalog reduces to two kernel classes, the two quadratures of the product of the forward and adjoint fields.
Real-valued temporal (broadband) signals are no exception, since every frequency component of a real waveform is still complex.
Therefore, every kernel in the catalog can be obtained as a linear combination of the two projections, which we label even and odd, the names taken from the ballistic limit.
The quadrature is selected either by a component of the material parameter or by the misfit, while seismology uses both.
This two-class structure originates from the complex field product rather than from a particular parametrization or ray geometry. Therefore, it persists for the coherent-wave kernel in the diffusive regime of, e.g., tissue optics, where multiple scattering smears any ray-based pattern.
We refer to the resulting unified treatment as coherent wave tomography (CWT), and in the Supplementary Notes we compare the terminologies to show the shared kernel structure explicitly in each area.

Second, we construct a phase diagram that places the five CWT areas on one plot, where two axes represent dimensionless quantities that determine wave propagation in a scattering medium~\cite{1999_vanRossum_multiple_scattering,2007_Akkermans_mesoscopic_physics,2012_Sato_Fehler_Maeda}.
The first is $k\ell_s$, the scattering mean free path $\ell_s$ in units of the reduced wavelength $\lambda/2\pi$ ($k=2\pi/\lambda$, in the medium), and the second is $L/\ell^*$, the source-to-detector path in units of the transport mean free path.
The diagram shows that full-wave inversions in the established areas independently operate within comparable limits, set by available high-performance computing (HPC) hardware.
These limits reflect a shared computational bottleneck~\cite{1996_Jo_optimal_9point,2007_Operto_3DFDFD,2016_Osnabrugge_bornseries,2007_Symes_checkpointing,2016_Komatitsch_anelastic_kernels_storage}, the storage of the wavelength-scale wavefields required to obtain the kernel.
The resources needed to compute and store them scale as $(L/\lambda)^3$ in three dimensions, for each retained frequency or time snapshot.

Third, we propose a wavefield compression scheme suitable for the broadband, temporal signals of FWI, ultrasound, GPR, and TomoSAR, so that one remedy serves all five areas of the diagram.
Compression has so far been used to alleviate the storage side of the bottleneck~\cite{2016_Boehm_wavefield_compression,2022_Kukreja_ZFP_compression,2023_Wang_Tucker_compression}, whereas the recently proposed On-Shell Compression And Reconstruction (OSCAR)~\cite{2026_Jara_OSCAR_compression} goes a step further by evaluating the kernel directly on the compressed monochromatic wavefield.
Because its applicability is governed by the $k\ell_s\gg 1$ criterion alone and not by the transport depth $L/\ell^*$, OSCAR covers both the quasi-ballistic and the diffusive regimes~\cite{2026_Jara_OSCAR_compression}.
Its broadband extension proposed here, OSCAR-BB, brings on-shell reconstruction to the time-domain areas.
We verify OSCAR-BB numerically in two and three dimensions over large bandwidths, reaching the broadband regime in which the time-domain areas operate.

Mitigation of the wavefield storage limitation can allow larger three-dimensional inversions in these areas, while the common compression scheme is an example of synergy enabled by the unified CWT approach.
\section{Results\label{sec:results}}

We first introduce the CWT description of one generic, area-neutral kernel and its two projections, for both monochromatic and broadband interrogation, Secs.~II\,A,~II\,B.

\begin{figure*}[t]
\centering
\includegraphics[width=0.8\textwidth]{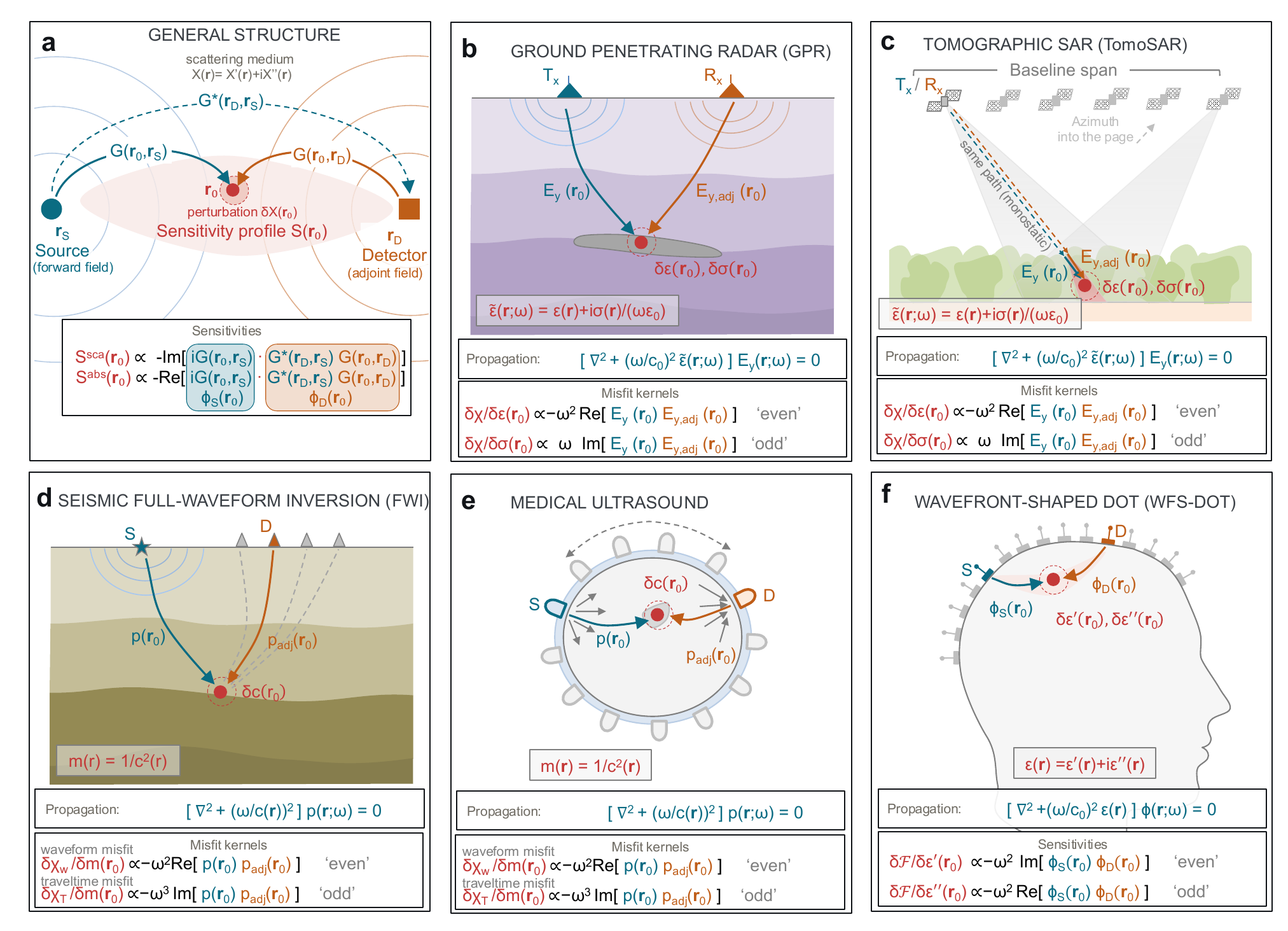} %
\caption{\label{fig:adjoint}%
Two-class kernel structure traced to the quadratures of the forward-adjoint product.
(a)~Generic structure, cf.~Secs.~II\,A,~II\,B. A perturbation $\delta X(\vec r_0)$ of the complex material parameter $X(\vec r)=X'(\vec r)+iX''(\vec r)$ contributes to the boundary signal via the product $\phi_\mathrm{S}(\vec r_0)\,\phi_\mathrm{D}(\vec r_0)$.
The two components of $\delta X$ enter through the two real-valued projections of this product, the sensitivities $S^{\mathrm{sca}}, S^{\mathrm{abs}}$ of Eq.~(3).
$\phi_\mathrm{S}$ is the forward field from source $\vec r_\mathrm{S}$, and $\phi_\mathrm{D}$ the adjoint field back-propagated from the conjugated detector signal at $\vec r_\mathrm{D}$.
(b--f)~Area-specific cases of (a), each with its wave equation and its kernel pair, one member from each class.
(b)~GPR: surface antennas, directed out of the survey plane, probe $\varepsilon(\vec r), \sigma(\vec r)$ of the subsurface. One representative transmitter-receiver pair is shown.
(c)~TomoSAR: cross-track baselines synthesize an elevation aperture, recovering $\varepsilon(\vec r)$ per range-azimuth cell in canopies and the near-surface.
One representative baseline ray is shown, while the gradient is the coherent sum over baselines.
(d)~Seismic FWI: a surface source-receiver array images $c(\vec r)$ at depth in a layered earth.
(e)~Medical ultrasound: an encircling (ring) transducer array images $c(\vec r)$ in breast tissue.
(f)~WFS-DOT: scalp fibers are envisioned to interrogate $\varepsilon'(\vec r), \varepsilon''(\vec r)$ in brain tissue via deterministic, sample-dependent near-infrared illumination.
In (b--e) the pair appears as the data-misfit (Fr\'echet) kernels, of the same product form, with the residual carried by the adjoint field.
In (b,~c) the two members are indexed by the parameter component, $\delta\chi/\delta\varepsilon$ vs $\delta\chi/\delta\sigma$, and in (d,~e) by the misfit, $\delta\chi_w/\delta m$ vs $\delta\chi_T/\delta m$.
In (f) the pair appears as the sensitivities of the observable, the scattering $\delta\mathcal{F}/\delta\varepsilon'$ (`even') and the absorption $\delta\mathcal{F}/\delta\varepsilon''$ (`odd').
}
\end{figure*}

\subsection{Monochromatic sensitivity kernels\label{sec:gxg}}
CWT recovers the spatial distribution of a material parameter $X(\vec r)$ from boundary measurements of the flux at a detector $\mathcal{F}[\vec r_\mathrm{D},\vec r_\mathrm{S}; X]$, where $X(\vec r)$ is in general complex.
In the Born approximation, the linear response of $\mathcal{F}$ to a localized perturbation $\delta X(\vec r_0)$ defines the sensitivity kernel
\begin{equation}\label{eq:sens_def}
S(\vec r_0) \;\equiv\; \frac{\delta\mathcal{F}[\vec r_\mathrm{D},\vec r_\mathrm{S}; X]}{\delta X(\vec r_0)}.
\end{equation}
Instead of a raster scan requiring a separate simulation for each $\vec r_0$, the kernel $S(\vec r_0)$ can be computed throughout the volume using only the two simulations of the forward and adjoint fields, as illustrated by the back-of-the-envelope argument below.

\emph{Setup:} At a fixed frequency $\omega$, the unperturbed field $\phi(\vec r)$ obeys the Helmholtz-like equation $[\nabla^2 + (\omega/c_0)^2\,X(\vec r)]\,\phi(\vec r) = 0$, with $c_0$ a constant reference wave speed and $\phi$ an area-neutral scalar field symbol adapted to specific wave-physics conventions in Sec.~II\,C, see also Supplementary Notes~S1, S2 and S3--S7 for full detail.
The Helmholtz-like form is adopted for concreteness.
The two-class structure derived below rests only on the field obeying a linear wave equation in which the material enters through a parameter at each point, so it carries over to the vector, elastic, and acoustic forms of Sec.~II\,C.
We model source and detector as waveguide ports supporting flux-normalized propagating modes $\{|\psi_a\rangle\}_{a=1}^{N_\mathrm{S}}$ and $\{|\psi_b\rangle\}_{b=1}^{N_\mathrm{D}}$~\cite{2022_Bender_coherent_remission,2025_Jara_coherent_wave_sensing}.
An illumination $\phi_\mathrm{S}(\vec r) = \sum_a v_a\,\phi_a(\vec r)$, with $v_a$ the input-mode amplitudes, excites the channel field $\phi_a$ from input port $a$, and the detected flux is $\mathcal{F} = \sum_b |u_b|^2$ with $u_b = \langle\psi_b|\phi_\mathrm{S}\rangle$ the unperturbed amplitude in output channel $b$.

\emph{Argument:} A pointlike perturbation $\delta X$ at $\vec r_0$ of effective volume $\Delta V$ produces, in the Born approximation, a scattered field $\delta\phi(\vec r) = \mathcal{V}\,G(\vec r,\vec r_0)\,\phi_\mathrm{S}(\vec r_0)$ with $\mathcal{V} \equiv -(\omega/c_0)^2\,\delta X\,\Delta V$ and $G$ the retarded Green's function~\cite{2025_Jara_coherent_wave_sensing}.
Linearization of $\mathcal{F}$ in $\delta\phi$ yields $\delta\mathcal{F}\approx\sum_b 2\,\mathrm{Re}\!\left[u_b^{*}\,\delta\phi_b\right]$, with $\delta\phi_b = \langle\psi_b|\delta\phi\rangle$ the change of the output amplitude.
Using the Fisher--Lee relation~\cite{1981_Fisher_Lee}, this collapses to a product of two fields evaluated at the perturbation point $\vec r_0$,
\begin{equation}\label{eq:dF_micro}
\delta\mathcal{F} = \mathrm{Re}\!\left[-i\,\mathcal{V}\,\phi_\mathrm{S}(\vec r_0)\,\phi_\mathrm{D}(\vec r_0)\right],
\quad
\phi_\mathrm{D}(\vec r_0) \equiv \sum_{b=1}^{N_\mathrm{D}} u_b^{*}\,\phi_b(\vec r_0).
\end{equation}
Here $\phi_b(\vec r_0)$ is the field at $\vec r_0$ produced by exciting only channel $b$ from the detector port, so that $\phi_\mathrm{D}(\vec r_0)$ is the field at $\vec r_0$ from back-propagating the conjugated detected amplitudes through the channel modes~\cite{2025_Jara_coherent_wave_sensing}, see Supplementary Note~S1.

\emph{Two modalities:} Let $X(\vec r) = X'(\vec r) + i\,X''(\vec r)$.
$X'$ sets the wave speed and scatters the field, whereas $X''$ is responsible for loss.
These two modalities yield two real-valued sensitivities,
\begin{equation}\label{eq:kernel_fields}
\begin{aligned}
S^{\mathrm{sca}}(\vec r_0) \;&\equiv\; \frac{\delta\mathcal{F}}{\delta X'(\vec r_0)} \;=\; -\frac{\omega^2}{c_0^2}\,\Delta V\,\mathrm{Im}[\phi_\mathrm{S}(\vec r_0)\,\phi_\mathrm{D}(\vec r_0)],\\
S^{\mathrm{abs}}(\vec r_0) \;&\equiv\; \frac{\delta\mathcal{F}}{\delta X''(\vec r_0)} \;=\; -\frac{\omega^2}{c_0^2}\,\Delta V\,\mathrm{Re}[\phi_\mathrm{S}(\vec r_0)\,\phi_\mathrm{D}(\vec r_0)].
\end{aligned}
\end{equation}
Equation~(3) is exact within the Born approximation and makes no assumption about transport regime, degree of disorder, or input-wavefront structure.

\emph{Point excitation/detection:} An analog of Eq.~(3) follows directly in terms of point-to-point Green's functions.
With $\phi_\mathrm{S}(\vec r_0) \propto i\,G(\vec r_0,\vec r_\mathrm{S})$ and $\phi_\mathrm{D}(\vec r_0) \propto (-i)\,G^{*}(\vec r_\mathrm{D},\vec r_\mathrm{S})\,(i)G(\vec r_\mathrm{D},\vec r_0)$, with one Fisher--Lee factor $i$ per Green's function and its conjugate in $u_b^{*}$, cf.~Supplementary Note~S1, and using spatial reciprocity $G(\vec r_\mathrm{D},\vec r_0) = G(\vec r_0,\vec r_\mathrm{D})$, Eq.~(3) reduces to
\begin{equation}\label{eq:kernel_G3}
S^{\mathrm{abs}}(\vec r_0) \;\propto\; -\,\mathrm{Re}\!\left[i\,G(\vec r_0,\vec r_\mathrm{S})\,G^{*}(\vec r_\mathrm{D},\vec r_\mathrm{S})\,G(\vec r_0,\vec r_\mathrm{D})\right]
\end{equation}
of Fig.~1(a), with the corresponding $\mathrm{Im}$-side expression for $S^{\mathrm{sca}}$.
The change of setup, port-mode versus point-source, contributes an overall factor of $i$ between Eq.~(3) and Eq.~(4).
The parameter-side labels $S^{\mathrm{sca}}$, $S^{\mathrm{abs}}$ are invariant under this rotation, while the kernel projection, $\mathrm{Re}$ or $\mathrm{Im}$ of the product of three Green's functions, is not.

\subsection{Broadband sensitivity kernels\label{sec:broadband}}

\begin{figure*}[t]
\centering
\includegraphics[width=\textwidth]{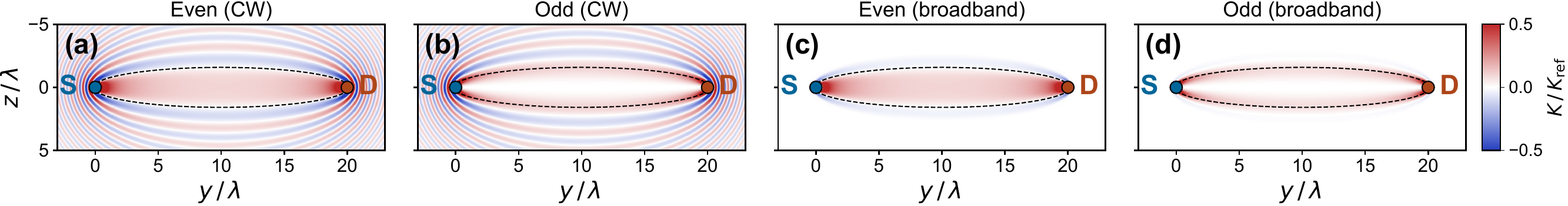} %
\caption{\label{fig:bananas}The `even' and `odd' kernel classes in an unbounded homogeneous (ballistic) three-dimensional (3D) background emerge as the two quadratures of one kernel, $\mathcal{P}(\vec r_0;\omega)$ of Eq.~(6).
(a, b)~Monochromatic (continuous-wave, CW) kernels, with Fresnel-zone oscillations at half-wavelength spacing.
(c, d)~Broadband kernels for a Gaussian spectral envelope of relative width $\sigma_\omega/\omega_0 = 0.5$.
In seismology, (c) and (d) are known as the waveform and the `banana--doughnut' traveltime kernels~\cite{1984_Tarantola_FWI,1999_Marquering_banana_doughnut,2000_Dahlen_banana_doughnut,2002_Dahlen_Baig_amplitude}.
Source S and detector D are separated by $20\lambda$.
The dashed curve is the excess-path-length contour $\delta L = \lambda/4$, the first zero of the CW even kernel.
Color scale $K/K_\mathrm{ref}$, with $K_\mathrm{ref}$ the peak of (c), is limited to $\pm0.5$, see Methods.}
\end{figure*}

The CWT formulation of Sec.~II\,A extends to broadband (BB) pulsed interrogation.
Its unperturbed forward field $\phi_\mathrm{S}(\vec r;t)$ now obeys the homogeneous time-domain wave equation.
The same perturbation $\delta X(\vec r_0)$, assumed here to be stationary, acts as a secondary time-dependent source $(\delta X\,\Delta V/c_0^2)\,\partial_{t'}^2\phi_\mathrm{S}(\vec r_0;t')$, with $\partial_{t'}^2$ replacing the monochromatic $-\omega^2$ factor of Sec.~II\,A.
Variation of the time-integrated detected flux $\mathcal{F}_\mathrm{BB} = \int_{-\infty}^{\infty} dt\sum_b |u_b(t)|^2$, analogous to Eq.~(2), gives the broadband absorption sensitivity
\begin{equation}\label{eq:imaging_condition}
S^{\mathrm{abs}}_\mathrm{BB}(\vec r_0) \;=\; -\frac{\Delta V}{c_0^2}\,\mathrm{Re}\!\int_{-\infty}^{\infty} dt'\;\partial_{t'}\phi_\mathrm{S}(\vec r_0;t')\,\partial_{t'}\phi_\mathrm{D}(\vec r_0;t'),
\end{equation}
cf.~Supplementary Note~S2, the broadband analog of Eq.~(3).
For the complex fields, the scattering counterpart $S^{\mathrm{sca}}_\mathrm{BB}$ takes $\mathrm{Im}$ of the same integrand, cf.~Supplementary Note~S2.
$\phi_\mathrm{D}(\vec r_0;t')$ is the channel-summed conjugated detector signal back-propagated via channel modes.
The sensitivity kernel is the product of the two fields at $\vec r_0$ evaluated at the same instant, integrated over time.
This is Claerbout's imaging condition of reflection seismology~\cite{1971_Claerbout_imaging}.

Parseval's theorem transforms the time pairing of the fields in Eq.~(5) into a per-$\omega$ pairing weighted by the source power spectrum $|j_\mathrm{s}(\omega)|^2$, with $j_\mathrm{s}(t)$ the emitted source pulse.
In the point-source/detector setup, the integrand reduces to the product of three Green's functions, cf.~Eq.~(4),
$\mathcal{P}(\vec r_0;\omega) \equiv G(\vec r_0,\vec r_\mathrm{S};\omega)\,G^{*}(\vec r_\mathrm{D},\vec r_\mathrm{S};\omega)\,G(\vec r_0,\vec r_\mathrm{D};\omega)$, with the Fisher--Lee factor $i$ excluded from the definition of $\mathcal{P}$.
Its real and imaginary parts are the even and the odd kernel classes introduced above, cf.~Fig.~2.
Accounting for the $\omega^2$ factor from the two time derivatives,
\begin{equation}\label{eq:broadband}
S^{\mathrm{abs}}_\mathrm{BB}(\vec r_0) \;\propto\; \int_{0}^{\infty}\! d\omega\;\omega^2\,|j_\mathrm{s}(\omega)|^2\,\mathrm{Im}[\mathcal{P}(\vec r_0;\omega)],
\end{equation}
the scattering counterpart taking the form of $-\mathrm{Re}$, cf.~Supplementary Note~S2.
In the quasi-ballistic regime, at fixed $\omega$, $\mathcal{P}(\vec r_0;\omega)$ oscillates in space, cf.~Fig.~2(a,~b).
Finite bandwidth $\Delta\omega$ dephases these oscillations away from the ballistic trajectory at excess path length $\delta L \gtrsim c/\Delta\omega$, the path via $\vec r_0$ in excess of the direct source-detector path, with $c$ the propagation speed in the background medium, cf.~Fig.~2(c,~d).
The unified description of Secs.~II\,A,~II\,B is applied to specific areas in the next section.

\Needspace*{4\baselineskip}
\subsection{Coherent wave tomography: \\
area-specific applications\label{sec:areas}}

\emph{GPR.} Ground-penetrating radar probes the shallow subsurface with broadband electromagnetic pulses at $10$~MHz--$1$\,GHz, recovering permittivity and conductivity by full-waveform inversion~\cite{2005_Annan_GPR,2007_Ernst_GPR_FWI,2011_Meles_GPR_FWI,2014_Lavoue_GPR_FWI,2019_Klotzsche_GPR_FWI_review}, see Fig.~1(b).
We adopt the two-dimensional (2D) description common in GPR FWI~\cite{2014_Lavoue_GPR_FWI} because the medium, to a good approximation, is invariant along the horizontal $y$-axis, the model plane is $(x,z)$ with $z$ the depth, and the antennas point along $y$.
In this configuration, known in the area as transverse electric (TE), the electric field retains a single out-of-plane component $E_y(\vec r;t)$, which behaves as a scalar and obeys a homogeneous wave equation with two real material parameters. The (real) permittivity $\varepsilon(\vec r)$ couples through $\partial_t^2 E_y$, and the conductivity $\sigma(\vec r)$ through $\partial_t E_y$, see Supplementary Note~S3.
In the crosshole geometry, where borehole dipoles excite the in-plane polarization, inversion proceeds through a scalar approximation~\cite{2007_Ernst_GPR_FWI} or the full 2D vector field~\cite{2010_Meles_GPR_vector}, with the kernel pair unchanged.
In the spectral domain the two parameters combine into a single complex effective permittivity $\tilde\varepsilon(\vec r;\omega) = \varepsilon(\vec r) + i\sigma(\vec r)/(\omega\varepsilon_0)$~\cite{1998_Jackson_em}, with $\varepsilon_0$ the vacuum permittivity, and the wave equation reduces to the generic Helmholtz-like form of Sec.~II\,A with $X(\vec r;\omega) = \tilde\varepsilon(\vec r;\omega)$.
GPR inversion minimizes a data-misfit functional $\chi$ defined as the squared $L_2$ residual between modeled and recorded fields, see Methods and Supplementary Note~S3.
The area-specific Fr\'echet kernels $\delta\chi/\delta\varepsilon$ and $\delta\chi/\delta\sigma$ form a dual pair, the material-parameter-side projections of Eq.~(6): $\delta\chi/\delta\varepsilon \propto -\omega^2\,\mathrm{Re}[\mathcal{P}]$, the even class of Fig.~2(c), and $\delta\chi/\delta\sigma \propto +\omega\,\mathrm{Im}[\mathcal{P}]$, the odd class of Fig.~2(d).
The lower-order $\partial_t$ coupling of $\sigma$ accounts for both the first power of $\omega$ and the factor $i$ that rotates one projection into the other, see Fig.~1.
GPR operates in the quasi-ballistic regime, see Fig.~3.

\emph{Tomographic SAR.} TomoSAR recovers the volumetric reflectivity of forest canopies and the near-surface from a sequence of repeat-pass synthetic-aperture radar (SAR) acquisitions taken at slightly offset cross-track positions (baselines)~\cite{2000_Reigber_Moreira_TomoSAR,2012_Tebaldini_Rocca_TomoSAR,2019_Lavalle_TomoSAR}, see Fig.~1(c).
Similar to GPR, the wave equation describes a single real-valued time-dependent TE component $E_y(\vec r;t)$.
The out-of-plane $y$-axis is directed along the azimuth direction, so the scalar description addresses the horizontally polarized channel of the acquisition, while material parameter components take the form of permittivity $\varepsilon(\vec r)$ and conductivity $\sigma(\vec r)$.
The even/odd kernel pair of Eq.~(3) applies unchanged, see Supplementary Note~S4.
Spectral estimation over the baselines then gives the reflectivity profile along elevation at each range-azimuth pixel~\cite{2000_Reigber_Moreira_TomoSAR,2012_Tebaldini_Rocca_TomoSAR}.
In a repeat-pass stack each acquisition is monostatic, so the per-baseline Fr\'echet kernel reduces to the GPR form with $\vec r_\mathrm{S}=\vec r_\mathrm{D}$ at the platform position.
Forest canopies are optically thin, which places TomoSAR in the quasi-ballistic regime, see Fig.~3 and Methods.
The multi-baseline inversion used at present recovers only the even member of the pair.
The odd member, the conductivity, should also be recoverable under full-wave processing, similar to GPR.

\emph{Seismic FWI and medical ultrasound.} Seismic FWI recovers subsurface wave-speed structure from broadband ground-motion records~\cite{1984_Tarantola_FWI,2006_Plessix_adjoint_state_review,2009_Virieux_FWI_review,2011_Fichtner_FWI_book}, see Fig.~1(d).
Medical ultrasound applies the same FWI to soft-tissue acoustic structure with transducer arrays~\cite{2015_Sandhu_ultrasound_FWI,2019_Wiskin_breast_ultrasound,2020_Guasch_FWI_brain,2022_Lucka_US_FWI}, see Fig.~1(e).
The acoustic pressure field $p(\vec r;t)$ is real and obeys a homogeneous wave equation with a single real material parameter $m(\vec r) = c^{-2}(\vec r)$, the scalar acoustic reduction of the elastic problem adopted here, see Methods and Supplementary Notes~S5 and~S6.
Identifying $X(\vec r) = (c_0/c(\vec r))^2$ in Sec.~II\,A gives $X''(\vec r) = 0$ in the common lossless treatment, cf.~Eq.~(3).
Inverting for loss restores the imaginary component~\cite{2013_Kamei_Pratt_viscoacoustic_FWI,2005_Tromp_adjoint_banana_doughnut,2019_Wiskin_breast_ultrasound,2009_Mulder_Hak_attenuation_ambiguity}.
Kernels of both classes can nevertheless be identified when $X''(\vec r) = 0$, via a waveform $L_2$ misfit~\cite{1984_Tarantola_FWI} and a cross-correlation traveltime misfit~\cite{1999_Marquering_banana_doughnut,2000_Dahlen_banana_doughnut,2005_Tromp_adjoint_banana_doughnut}.
The gradients of the two misfits, $\chi_w$ and $\chi_T$, with respect to the same parameter $m(\vec r)$ differ by one $\partial_t$ due to a difference in the adjoint sources.
This extra time derivative corresponds to a factor $-i\omega$.
The two gradients become $\delta\chi_w/\delta m \propto -\omega^2\,\mathrm{Re}[\mathcal{P}]$, the even class of Fig.~2(c), and $\delta\chi_T/\delta m \propto -\omega^3\,\mathrm{Im}[\mathcal{P}]$, the odd class of Fig.~2(d).
The latter is the canonical Fr\'echet kernel of finite-frequency tomography, the seismic `banana--doughnut'~\cite{1999_Marquering_banana_doughnut,2000_Dahlen_banana_doughnut,2004_Spetzler_Snieder_Fresnel}.
In broadband areas the imaginary projection also describes the arrival delay, through which FWI and ultrasound sense wave speed~\cite{1991_Luo_Schuster_traveltime,2005_Tromp_adjoint_banana_doughnut}.
The adjoint field of Eq.~(4) is the time-reversed field of acoustic time-reversal mirrors~\cite{1992_Fink_time_reversal_I,2000_Fink_time_reversal,1994_Prada_Fink_DORT}, related by the reciprocity relations underlying Green's-function retrieval~\cite{2004_Wapenaar_retrieval}.
Seismic FWI and ultrasound operate in the quasi-ballistic regime, see Fig.~3.
The crust and welded-tuff entries in Table~I are coda-envelope values, cf.~Fig.~3~\cite{2012_Sato_Fehler_Maeda,2009_Przybilla_Korn,2017_Eulenfeld_Wegler_USarray}.

\emph{WFS-DOT.} Wavefront-shaped diffuse optical tomography is a recent proposal~\cite{2022_Bender_coherent_remission,2025_Jara_coherent_wave_sensing,2026_Jara_OSCAR_compression} that aims to use monochromatic, deterministic, sample-dependent illumination of biological tissue at NIR wavelengths, see Fig.~1(f).
Within the WFS-DOT description, the material parameter is the complex permittivity $\varepsilon(\vec r) = \varepsilon'(\vec r) + i\varepsilon''(\vec r)$, identifying $X(\vec r) = \varepsilon(\vec r)$ in Sec.~II\,A with both components simultaneously present.
The Re/Im pair of Eq.~(3) is then a direct parameter splitting: $\delta\mathcal{F}/\delta\varepsilon'(\vec r_0) \propto -\omega^2\,\mathrm{Im}[\phi_\mathrm{S}\phi_\mathrm{D}]$, the refractive branch, of the even class, and $\delta\mathcal{F}/\delta\varepsilon''(\vec r_0) \propto -\omega^2\,\mathrm{Re}[\phi_\mathrm{S}\phi_\mathrm{D}]$, the absorption branch, of the odd class, see Fig.~1.
At the transport scale $\ell^*$, these wave-level parameters coarse-grain onto the diffuse observables, $\delta\varepsilon'' \leftrightarrow \delta\mu_a$ (absorption coefficient) while index contrast in $\varepsilon'$ determines the reduced scattering coefficient $\mu_s'$, see Supplementary Note~S7.
Conventional DOT operates on these coarse-grained quantities through a parallel incoherent description built on the diffusion equation and ensemble-averaged photon fluences~\cite{1993_Schotland_photon_hitting_density,1995_Arridge_PMDF,1995_Yodh_Chance_spectroscopy,1999_Arridge_optical_tomography,2010_Durduran_diffuse_optics,2004_Boas_Dale_DOT_imaging}, valid under random-input excitation and disorder averaging.
WFS-DOT differs from conventional DOT in that it {\it controls the coherent input field}, retaining its phase.
This is {\it distinct from the incoherent spatially structured intensity} and dense-source illumination already utilized in DOT~\cite{2009_Cuccia_SFDI,2014_Eggebrecht_HDDOT}.
WFS-DOT is envisioned to operate in the diffusive regime where the wave-level kernel is smeared by multiple scattering.
The two-class kernel structure is independent of the transport regime and forces the even/odd split here as well, where it cannot be reduced to the cosine/sine dependence on the excess path seen in the quasi-ballistic case.
WFS-DOT is proposed to operate with both $k\ell_s$ and $L/\ell^*$ simultaneously large, see Fig.~3.

\subsection{Phase diagram\label{sec:kernels}}

\begin{figure*}[!htbp]
\begin{minipage}[c]{0.70\textwidth}
\centering
\includegraphics[width=\textwidth]{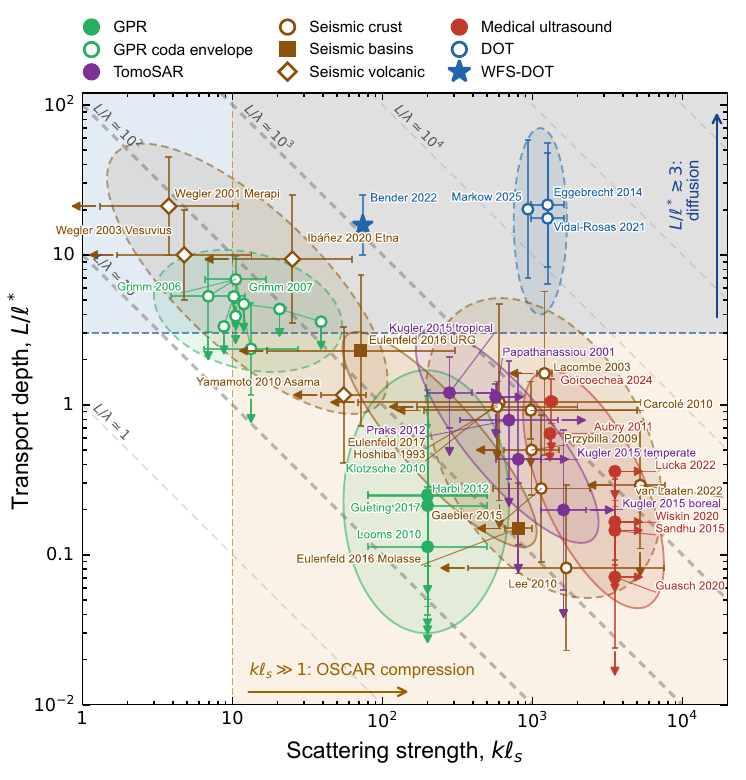}
\end{minipage}\hfill
\begin{minipage}[c]{0.28\textwidth}
\caption{\label{fig:parameter_space}%
Phase diagram of the CWT areas.
Each point represents a published study placed according to the reported or deduced scattering strength $k\ell_s$ and the transport depth $L/\ell^*$, see Methods.
Bars show the range of a parameter in the study.
Arrows mark bounds, with the value lying in the arrow direction.
Lines join data sets, e.g., a frequency sweep, in the same study.
Filled symbols denote inversions of the coherent wave field, open symbols inversions of intensities or envelopes.
Pale ovals outline the area groups, solid for field-based and dashed for intensity-based inversions.
The star is the wavefront-shaping experiment behind the WFS-DOT proposal~\cite{2022_Bender_coherent_remission}.
Orange band marks the regime of OSCAR compression, $k\ell_s\gg1$~\cite{2026_Jara_OSCAR_compression}.
Blue band marks the onset of diffusive transport, $L/\ell^*\gtrsim3$, see Supplementary Note~S7.
Gray dashed lines are contours of constant $L/\lambda$ under the $\ell^*\approx2\pi\ell_s$ assumption, see Methods.}
\end{minipage}
\end{figure*}

The unified CWT approach of Secs.~II\,A--II\,C enables analysis of the area-specific applications in terms of two quantities that characterize wave transport, cf.~Fig.~3.
$k\ell_s$ plotted on the horizontal axis is the scattering strength per wavelength, with $\ell_s$ the scattering mean free path~\cite{2007_Akkermans_mesoscopic_physics,1999_vanRossum_multiple_scattering}.
$L/\ell^*$ plotted on the vertical axis is the cumulative scattering over the source-to-detector path $L$, with $\ell^* = \ell_s/(1-g)$ the transport mean free path~\cite{1998_Margerin_radiative_transfer,2012_Sato_Fehler_Maeda,2007_Akkermans_mesoscopic_physics}, see Methods.
Diagonal contours of constant $L/\lambda$ have slope $-1$ on a log-log scale under the $\ell^*\approx 2\pi\ell_s$ assumption, see Methods.
These contours anchor levels of computational cost $\sim (L/\lambda)^3$ based on the same dimensionless ratio for all areas, requiring a common conversion rather than granular per-area assumptions on the anisotropy $g$.

The CWT studies of Fig.~3, the filled symbols, cluster within the band $1 \lesssim L/\lambda \lesssim 10^3$, despite spanning media as different as soft tissue, sand and gravel aquifers, forest canopies, and sedimentary basins.
The open symbols are inversions of \emph{intensities or envelopes} that include seismic coda in the crust and under volcanoes, GPR coda in welded tuff, and diffuse light in tissue.
They depict media where the transport parameters are known but \emph{coherent inversion} has not been implemented.
The upper edge of the $L/\lambda$ band is set by computational tractability, which becomes evident from the following argument.
Full-waveform inversions run on models of $10^7$--$10^8$ voxels~\cite{2007_Operto_3DFDFD,2013_Warner_3D_FWI}, while single forward simulations reach $10^9$ voxels and beyond~\cite{2003_Komatitsch_EarthSimulator}.
This places wave simulations resolving $(L/\lambda)^3 \sim 10^7$ wavelength-scale cells at the limit of current HPC hardware.
Within the band, adjoint-based full-wave inversion is established practice in GPR and seismic FWI, and is emerging in medical ultrasound.

WFS-DOT aims to bring wave-level tomography, and the sensitivity gain it promises, to tissue~\cite{2025_Jara_coherent_wave_sensing}.
At NIR wavelengths this area would sit one to two orders past the quasi-ballistic cluster, where DOT operates in Fig.~3, at $L/\lambda \sim$ $10^4$--$10^5$.
Experimentally, only a proof-of-concept demonstration on a specially designed platform exists at this point~\cite{2022_Bender_coherent_remission}, the star in Fig.~3.
Reconstruction is further confounded by the fact that the wave-level kernel is smeared by multiple scattering~\cite{2025_Jara_coherent_wave_sensing}.
Conventional DOT side-steps these difficulties at the cost of resolution and phase information, see Supplementary Note~S7.
The $(L/\lambda)^3 \sim$ $10^{12}$--$10^{15}$ wavefield storage/computation cost places WFS-DOT well beyond current HPC reach.

Furthermore, in living tissue, physiological processes, predominantly blood flow and Brownian motion of the scatterers, continuously perturb the interference pattern.
At millimeter depths in the living brain the speckle pattern decorrelates in about a millisecond~\cite{2017_Qureshi_invivo_decorrelation}, and at centimeter-scale source-detector separations on the human head it decays within tens of microseconds~\cite{2021_Zhou_iDWS}.
Therefore, a coherent measurement must deliver its forward-and-adjoint pass within this time, which currently makes in-vivo applications beyond reach experimentally.

\subsection{On-shell structure and OSCAR compression\label{sec:oscar}}

\begin{figure}[!htbp]
\centering
\includegraphics[width=\columnwidth]{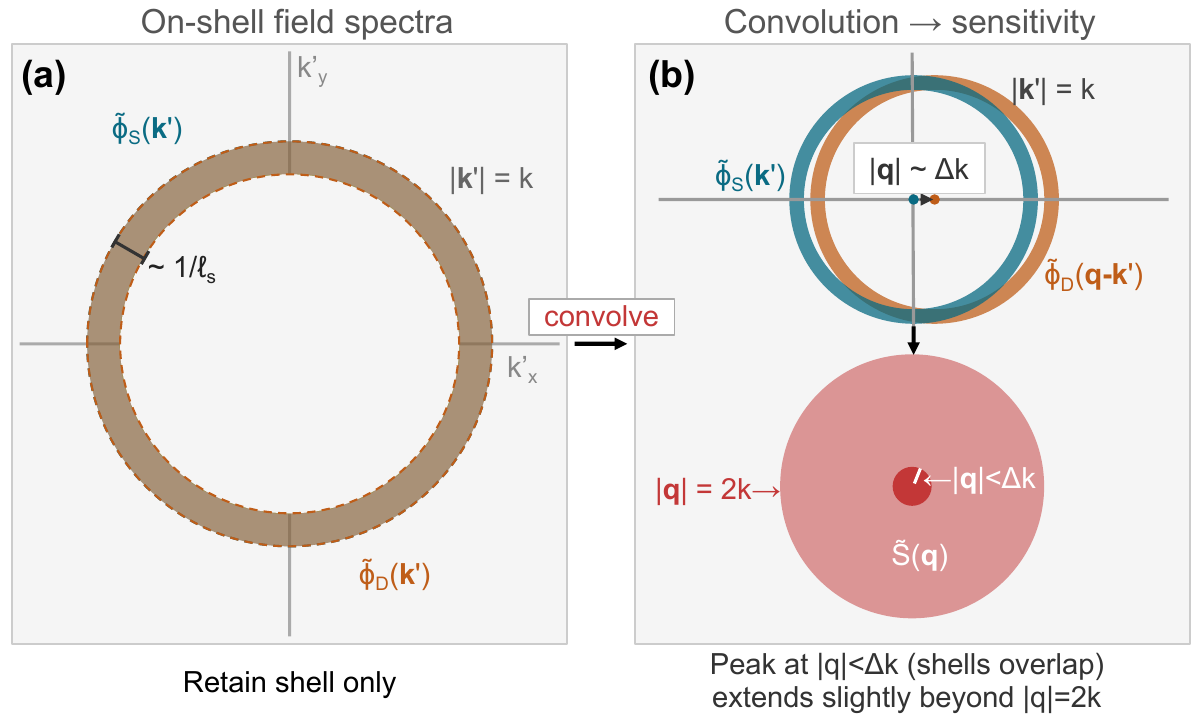} %
\vskip -0.2cm
\caption{\label{fig:oscar}On-shell structure underlying OSCAR compression~\cite{2026_Jara_OSCAR_compression}.
(a)~The spatial Fourier transforms of the forward field, $\widetilde{\phi}_\mathrm{S}(\vec{k}')$ (teal), and of the adjoint field, $\widetilde{\phi}_\mathrm{D}(\vec{k}')$ (orange), both concentrate on a thin shell of radius $|\vec{k}'|=k$ and thickness ${\sim}1/\ell_s$ in Fourier space, retained within a mask of width $\Delta k=\alpha/\ell_s$.
The two shells coincide, so only one ring is drawn to represent both fields, while the dashed outline distinguishes $\widetilde{\phi}_\mathrm{D}$.
Retaining only the on-shell data compresses each stored field.
(b)~Top: the sensitivity spectrum $\widetilde{S}(\vec{q})$, the convolution of the two field spectra, see Supplementary Note~S9, is the overlap of the two shells when one is shifted by $\vec{q}$.
For $|\vec{q}| \lesssim \Delta k$ the shells remain overlapping along the entire circumference, giving the peak of $\widetilde{S}$.
The overlap decreases as $|\vec{q}|$ grows and falls off beyond $|\vec{q}|\approx 2k$.
Bottom: the resulting map of $\widetilde{S}(\vec{q})$.
The entire computation can be performed on compressed shell data, with no decompression.
Both panels are conceptual illustrations, not renderings of numerical data.}
\end{figure}

Figure~3 exposed a shared computational bottleneck imposed by wavefield storage in CWT.
In the time-domain codes, several mitigation strategies are used to avoid storing the full time history of the forward field, among them checkpointing~\cite{2000_Griewank_revolve,2007_Symes_checkpointing} and reconstruction of the forward field from stored boundary values~\cite{2014_Yang_boundary_saving} combined with accumulation of the gradient during the backward pass.
The storage of fields becomes a bottleneck when these strategies do not apply or are insufficient, e.g., in frequency-domain multi-source operation~\cite{2007_Operto_3DFDFD} and in WFS-DOT, where every forward field of a large source set needs to be retained.
Beyond these generic strategies, each area has developed its own remedies, from optimized finite-difference stencils to lossy compression and dimensional reduction, see Supplementary Note~S9.

We observe that most of the areas lie in the $k\ell_s\gg 1$ regime, the orange band in Fig.~3.
This is the locally weakly scattering regime in which the field's spatial Fourier transform (FT) components are concentrated on a shell of radius $k$ and thickness $\sim 1/\ell_s$~\cite{1999_vanRossum_multiple_scattering,2007_Akkermans_mesoscopic_physics}.
A physics-based compression scheme, OSCAR~\cite{2026_Jara_OSCAR_compression}, has recently been proposed to exploit this on-shell structure.
OSCAR involves the following steps:
(i) construction of a mask that selects which spatial FT components are retained, an isotropic shell set by $k$ and $\ell_s$ alone for domains deeper than $2\pi\ell_s$.
(ii) sample-agnostic on-shell compression by discarding field components outside of the mask, cf.~Fig.~4(a).
(iii) sensitivity-kernel evaluation directly from compressed shell data, cf.~Fig.~4(b), with wavelength-scale resolution preserved by the convolution procedure.
(iv) optional coarse-graining of the output sensitivity to the scale at which the gradient is used, the model cell or smoothing length in the quasi-ballistic regime and the transport scale $\ell^*$ in the diffusive regime where the sensitivity is statistically smooth.

Step (iii) makes OSCAR more than a storage scheme. The sensitivity kernel can be computed by convolution directly in the compressed state, cf.~Fig.~4(b).
The convolution yields the complex product itself, so both kernel classes, even and odd, follow from the same compressed data at no extra cost.
Because the shell mask is the same for every realization of the medium, the compressed fields combine without the need to reconstruct them to the original spatial grid.
Generic compression methods, by contrast, only reduce storage, and each field must be reconstructed to the full spatial grid before the kernel can be formed.

The broadband, time-domain extension of the scheme, OSCAR-BB, is described in Supplementary Note~S10, with its two field compression protocols defined in Methods.
In the per-frequency protocol, one assumes that the field data are available for all times, so that the field at each frequency is obtained and compressed with the per-frequency mask.
By construction, the CW compression ratio applies, and the CW procedure requires only minor modifications, such as frequency tracking of the mask.
The per-snapshot protocol represents the case in which the entire time series is not available and the field is processed one snapshot at a time.
The spatial Fourier components of a snapshot fill the annulus that spans all shells of the band, and the whole annulus has to be retained, which reduces the compression ratio.

With the Nyquist factor $n_{\Delta x}^d/c_d$ common to all areas, the per-snapshot compression ratio becomes,
{
\begin{equation}\label{eq:OSCAR_ratio}
C_\mathrm{BB}\;\simeq\;\frac{n_{\Delta x}^d}{c_d}\,
\frac{k\ell_s/\alpha}{\,1+B\,k\ell_s/\alpha\,},
\qquad
B\equiv\frac{\Delta\omega}{\omega_0},
\end{equation}
}%
where $n_{\Delta x}\equiv\lambda/\Delta x$ is the number of grid points per wavelength, $d$ the spatial dimension, $c_d=2\pi$ ($4\pi$) in 2D (3D), $\alpha\equiv\Delta k\,\ell_s$ the full mask width in units of the shell width, and $B$ the normalized bandwidth, with $\omega_0$ the center frequency and $\Delta\omega$ the full spectral span.
A mask width $\alpha\approx 10$ gives a $3\%$ overlap error between the rebuilt and the original fields in the large-domain limit~\cite{2026_Jara_OSCAR_compression}.
The bandwidth enters through the single crossover factor $1/(1+x)$, with $x\equiv B\,k\ell_s/\alpha$ the band sweep, the shift of the shell radius across the band, measured in mask widths.
In the narrow-band operation, $B\ll\alpha/k\ell_s$, the ratio reduces to the CW value $C_\mathrm{CW}=k\ell_s\,n_{\Delta x}^d/(c_d\alpha)$.
In the broadband operation, $B\gg\alpha/k\ell_s$, the ratio approaches $(\omega_0/\Delta\omega)\,n_{\Delta x}^d/c_d$, independent of the shell width, since the retained radial extent, the radial factor, is then mostly set by the band sweep rather than by scattering.
The first factor in Eq.~(7), the Nyquist factor, stays modest, due to $n_{\Delta x}\approx 3$--$7$, consistent across the CWT areas.
Using the $k\ell_s$ in Fig.~3 and $n_{\Delta x}=4$--$5$ common to the frequency-domain solvers in GPR~\cite{2014_Lavoue_GPR_FWI,2019_Klotzsche_GPR_FWI_review}, TomoSAR~\cite{2000_Reigber_Moreira_TomoSAR,2012_Tebaldini_Rocca_TomoSAR}, seismic FWI~\cite{2007_Operto_3DFDFD,2022_Aghamiry_adaptive} and medical ultrasound~\cite{2015_Sandhu_ultrasound_FWI}, $C_\mathrm{CW}^\mathrm{3D}$ at $\alpha=10$ reaches ${\sim}$$5{\times}10^2$ in GPR, ${\sim}$$1.2{\times}10^3$--$1.5{\times}10^3$ in medical ultrasound, up to ${\sim}4{\times}10^3$ in the crust and, assuming full-wave processing, $10^2$--$2{\times}10^3$ in TomoSAR.
For WFS-DOT in tissue optics with $n_{\Delta x}=3$--$7$~\cite{2022_Lin_MESTI,2016_Osnabrugge_bornseries}, $C_\mathrm{CW}^\mathrm{3D}$ evaluates to ${\sim}2{\times}10^2$--$4{\times}10^3$.
Ref.~\cite{2026_Jara_OSCAR_compression} measured $C_\mathrm{CW}^\mathrm{3D}\approx 420$ at a $3\%$ overlap error in a tissue-like medium at $k\ell_s\approx 3000$, and Eq.~(7) reproduces this value within a few percent for the mask width of that test, see Supplementary Note~S9.
Supplementary Fig.~S4 verifies Eq.~(7) for frequency combs in 2D and 3D.

Serendipitously, the annulus also retains the shell tails discarded by the monochromatic masks in the per-frequency protocol.
As a result, the reconstruction errors of the pulse and of the sensitivity kernel \emph{decrease} with the bandwidth by the same factor $1/(1+x)$.
We quantify the fidelity of the reconstruction by the overlap error $\epsilon_\phi$ between the pulse rebuilt from the compressed data and the original pulse, Eq.~(16) of the Methods.
A mask of width $\alpha$ discards the fraction $P_\mathrm{out}(\alpha)\simeq2/(\pi\alpha)=\bar P_{\mathrm F}$ of the power of a monochromatic Lorentzian shell~\cite{2026_Jara_OSCAR_compression}.
To leading order the per-frequency error is half of the discarded fraction averaged over the band, $\epsilon_\phi^{\mathrm F}\simeq\bar P_{\mathrm F}/2$, independent of the bandwidth.
The annulus of the per-snapshot protocol is wider by the factor $1+x$ and discards proportionally less power $\bar P_{\mathrm S}$, so that
\be\label{eq:SI_overlap_prediction}
\frac{\bar P_{\mathrm S}}{\bar P_{\mathrm F}}\simeq\frac{1}{1+x}\quad\Rightarrow\quad
\epsilon_\phi^{\mathrm{S}}(x)\simeq\frac{\epsilon_\phi^{\mathrm F}}{1+x},
\ee
with no adjustable parameters, see Supplementary Note~S10 for the derivation.
Figure~5 confirms both statements numerically for different systems for both 2D scalar and 3D vector waves, see Methods for simulation details.

In the per-frequency protocol the error stays within about $1\%$ of its narrow-band value over almost two decades of $x$, while in the per-snapshot protocol it follows Eq.~(8) above the crossover at $x\approx1$.
The error of the coarse-grained sensitivity kernel decreases in the same proportion, by a factor of $2.6$ to $26$ at the largest simulated $x$ and a cell size of four vacuum wavelengths, see Supplementary Note~S10.
\begin{figure}[!htbp]
\centering
\includegraphics[width=\columnwidth]{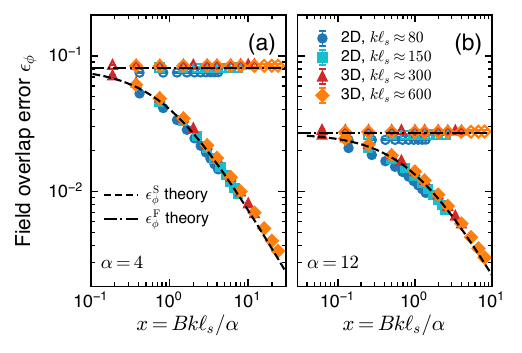} %
\vskip -0.2cm
\caption{\label{fig:oscarbb_error}Fidelity of the pulse reconstruction vs bandwidth.
Field overlap error $\epsilon_\phi$ over the stored volume, Eq.~(16) of the Methods, vs the band-sweep parameter $x=Bk\ell_s/\alpha$ in the per-frequency protocol (open symbols) and the per-snapshot protocol (filled symbols), in the four systems of Table~II, at $\alpha=4$ (a) and $12$ (b).
Symbols correspond to the average over an ensemble of $N_\mathrm{rlz}=10$ realizations and error bars to the standard deviation.
Black dash-dotted and dashed lines are theory predictions, see Supplementary Note~S10.}
\end{figure}

Application of OSCAR-BB can potentially advance the iso-$L/\lambda$ feasibility boundary in Fig.~3.
In a situation where wavefield storage is the dominant limitation, the stored voxel count drops by the factor $C_\mathrm{BB}$ and the accessible linear dimension of the domain grows by ${\sim}C_\mathrm{BB}^{1/3}$.
The validated $C_\mathrm{CW}\approx 420$ therefore translates into a factor of about $7.5$ in linear dimension, close to a full decade on the iso-$L/\lambda$ scale.
When one of the dimensions of the inversion domain is fixed, as in depth-limited geometries, the gain concentrates in the remaining ones.
Along a single free dimension, it can reach the full $C_\mathrm{BB}$.
In the proposed WFS-DOT and in frequency-domain seismic FWI the fields are stored one frequency at a time, so both lie in the narrow-band branch of Eq.~(7).
In WFS-DOT, where the large $k\ell_s\sim$$10^3$ enables substantial compression ratios, OSCAR may bring the prohibitive wavefield storage cost closer to HPC tractability.
In the broadband areas (GPR, time-domain FWI, broadband TomoSAR, and medical ultrasound), the same compression may reduce I/O across different source-detector combinations~\cite{2019_Klotzsche_GPR_FWI_review,2020_Guasch_FWI_brain,2022_Kukreja_ZFP_compression} as well as enable on-board inversion with a reduced downlink.

\section{Discussion\label{sec:discussion}}

We have formulated coherent wave tomography as one inverse problem common to GPR, TomoSAR, seismic FWI, medical ultrasound, and WFS-DOT.
The phase diagram of Fig.~3 places the areas on the shared transport axes $(k\ell_s,\,L/\ell^*)$, making the cross-area comparison quantitative and indicating where a method developed in one area may apply in another.
From the seismic side, the adjoint-state methods developed over four decades for heterogeneous and anisotropic media~\cite{1984_Tarantola_FWI,2009_Virieux_FWI_review,2006_Plessix_adjoint_state_review,2011_Fichtner_FWI_book} already underpin the established GPR and medical ultrasound inversions.
The same methods carry over at the kernel level to TomoSAR, once it is posed as a full-wave problem.
Full-wave processing would also recover the odd member of the pair, which would resolve the dissipative share of the canopy loss per voxel, whereas polarimetric interferometry retrieves the total extinction only as a volume average~\cite{2001_Papathanassiou_Cloude_RVoG,2012_Praks_extinction_Lband}.

These adjoint-state methods extend to WFS-DOT as well, although the diffusive transport in tissue complicates the inversion, cf.~Sec.~II\,D.
At sub-millimeter depths, wave-level optical imaging is already practiced as interferometric synthetic-aperture microscopy, demonstrated in the living human retina~\cite{2015_Shemonski_retina}.
Such microscopy methods evaluate the imaging condition, Eq.~(5), in a single pass, without residual weighting and model iteration. Therefore, they belong to migration and beamforming rather than to CWT, see Methods.

WFS-DOT suggests that sample-dependent input wavefronts can be optimized to enhance sensitivity~\cite{2021_Bouchet_max_information,2022_Bender_coherent_remission,2025_Jara_coherent_wave_sensing}.
Techniques tied to a specific transport regime, however, do not automatically apply across the CWT areas.
The maximum-remission eigenchannel of WFS-DOT~\cite{2022_Bender_coherent_remission,2025_Jara_coherent_wave_sensing} can be defined in any regime, but its properties should change drastically outside diffusive transport.
The beam-focusing excitation of seismic and ultrasound arrays similarly presumes quasi-ballistic propagation, so each requires a separate analysis.
Operating at the onset of diffusive transport, coda-wave interferometry~\cite{2002_Snieder_coda_interferometry,2003_Campillo_Paul} is the closest counterpart of WFS-DOT among the areas of Fig.~3.
It monitors change of the medium over time and localizes it with intensity-transport kernels~\cite{2005_Pacheco_Snieder_coda_sensitivity}.
Like conventional DOT, the envelope analyses of volcanic edifices, and the radar coda fits in welded tuff, this localization relies on intensity rather than on the coherent wavefield, which sets the dashed areas (open symbols) of Fig.~3 apart from coherent wave tomography.

Although the kernels of Sec.~II were obtained via linearization and are exact only in the Born approximation, CWT does recover large contrast via iteration from an adequate starting model~\cite{1984_Tarantola_FWI,2009_Virieux_FWI_review}.
At each step the gradient of the misfit is the sum over source-receiver pairs of the residual-weighted Fr\'echet kernels recomputed in the updated model, see Methods.
Each step inherits the two-class structure, as the gradients remain the two projections of the recomputed product $\phi_\mathrm{S}\phi_\mathrm{D}$.

Our common approach to the wave tomography can help to identify, e.g., constraints shared by different areas, the computational cost being one example considered in this work.
OSCAR~\cite{2026_Jara_OSCAR_compression} is proposed as a cross-area remedy, cf.~Fig.~4 and Sec.~II\,E.
The gain is largest in narrow-band or CW operation and more modest in broadband time-domain operation, where the radial factor is capped at $1/B$, see Eq.~(7) and Supplementary Fig.~S4.
Because OSCAR relies on $k\ell_s\gg 1$, it does not apply to locally strongly scattering systems such as volcanic edifices, where $k\ell_s$ can fall to $\sim 2$~\cite{2001_Wegler_Luhr_volcano}.

The two-class structure extends even beyond the CWT description, as the absorption sensitivity of conventional DOT is the diffusion limit of its absorption member, see Supplementary Note~S7.

\section*{Methods}

\subsection*{Minimization problem:\\ Misfit functional and the Fr\'echet kernel}

In GPR, seismic FWI, and medical ultrasound, wave-based tomography is conventionally set up~\cite{1984_Tarantola_FWI,2009_Virieux_FWI_review,2019_Klotzsche_GPR_FWI_review,2022_Lucka_US_FWI} as the minimization over the medium parameter $X(\vec r)$ of a scalar (error-estimate) misfit functional $\chi[X]$.
Improvement to $X(\vec r)$ on a successive iteration requires the functional gradient $\delta\chi/\delta X(\vec r_0)$ at every interior voxel.
The adjoint-state method~\cite{1984_Tarantola_FWI,2006_Plessix_adjoint_state_review} computes this gradient from one forward and one adjoint simulation. 
It shares the $\phi_\mathrm{S}(\vec r_0)\,\phi_\mathrm{D}(\vec r_0)$ structure with the sensitivity in Sec.~II\,A, Eq.~(3), except that the adjoint field now back-propagates the data residual and the contributions are summed over source-receiver pairs.
This gradient is referred to as the misfit kernel or the Fr\'echet kernel of the misfit functional~\cite{1999_Marquering_banana_doughnut,2000_Dahlen_banana_doughnut,2005_Tromp_adjoint_banana_doughnut,2004_Spetzler_Snieder_Fresnel}, and the kernels of Sec.~II\,C and Fig.~1 are its special cases, with one parameter-side branch for real $X(\vec r)$ and two for complex.

\emph{GPR and TomoSAR.} The misfit is the waveform $L_2$ residual of the TE field $E_y(\vec r_\mathrm{D};t)$.
The adjoint source is the data residual itself, back-propagated from the receiver.
One field pair produces two Fr\'echet kernels because the two components of the medium parameter enter the time-domain wave operator through different time derivatives.
The kernel $\delta\chi/\delta\varepsilon$ pairs the adjoint field with $\partial_{t'}^2 E_y$, and $\delta\chi/\delta\sigma$ with $\partial_{t'} E_y$, see Supplementary Note~S3, Eqs.~(S31)--(S33).
Under full-wave processing, which is not part of the currently common multi-baseline spectral estimation, the same misfit and Fr\'echet kernels would also apply to TomoSAR per acquisition, see Supplementary Note~S4.

\emph{Seismic FWI and medical ultrasound.} For the acoustic pressure field $p(\vec r;t)$ with model parameter $m(\vec r) = c^{-2}(\vec r)$, two of the commonly used misfit functionals exemplify the two kernel classes, the waveform $L_2$ residual $\chi_w$ and the cross-correlation traveltime residual $\chi_T$~\cite{1984_Tarantola_FWI,1999_Marquering_banana_doughnut,2000_Dahlen_banana_doughnut,2005_Tromp_adjoint_banana_doughnut}.
The two differ only in the quantity used as the adjoint source at the receiver, the data residual $r_w(t)$ for $\chi_w$ or $r_T(t) \propto -\partial_t p(\vec r_\mathrm{D};t)$ for $\chi_T$.
In practice, seismic FWI inverts the elastic wave equation, with additional coupling terms that do not change the structural form of the Fr\'echet kernel, see Supplementary Note~S5.
Back-propagating the chosen adjoint source and pairing it with $\partial_{t'}^2 p$ gives the Fr\'echet kernel $\delta\chi/\delta m$, see Supplementary Note~S5, Eqs.~(S38)--(S42).
Substituting $r_w$ for the adjoint source $r(t)$ gives the waveform kernel, the even class, and $r_T$ the traveltime kernel of finite-frequency tomography, the odd class, see Fig.~2(c,~d) and Fig.~1.

\subsection*{One-pass imaging techniques}

A single, unweighted application of the imaging condition, Eq.~(5), correlates the back-propagated data with the forward field.
It involves neither of the two steps that define the minimization above, the residual weighting and the iterative model update.
This distinct one-pass modality is readily exploited in each area.
Exploration seismology applies it as migration imaging~\cite{1971_Claerbout_imaging,1978_Stolt_migration}, medical ultrasound as delay-and-sum beamforming~\cite{2005_Szabo_diagnostic_ultrasound}, and SAR as aperture synthesis~\cite{1991_Curlander_McDonough_SAR,1999_Soumekh_SAR}.
In optics, the same principle is exploited in optical coherence tomography (OCT)~\cite{1991_Huang_OCT,2008_Drexler_retinal_OCT}.
Interferometric synthetic-aperture microscopy~\cite{2007_Ralston_ISAM} and computational adaptive optics~\cite{2012_Adie_CAO} refine OCT by solving the linearized (Born) inverse-scattering problem in closed form, the optical analog of Fourier-domain migration~\cite{1978_Stolt_migration}.
Such techniques, however, do not compute both forward and adjoint wavefields throughout the volume of the system.
Therefore, they are not subject to the $(L/\lambda)^3$ storage constraint that governs the areas in Fig.~3.
In contrast, the closest optical counterpart of FWI, an iterative adjoint-state (full-wave) optical inversion, has already been realized~\cite{2018_Liu_SEAGLE}, but has so far been applied as transmission diffraction tomography of nearly transparent systems rather than for backscatter imaging of tissue.

\subsection*{Even and odd kernels \\ in homogeneous background}

Figure~2 plots the $\mathrm{Re}$ and $\mathrm{Im}$ projections of the product of three Green's functions $\mathcal{P}$ of Eq.~(6) for an unbounded homogeneous background with wave speed $c$, $G(\vec r,\vec r') \propto e^{i(\omega/c)|\vec r-\vec r'|}/|\vec r-\vec r'|$.
The source at $\vec r_\mathrm{S}$ and the detector at $\vec r_\mathrm{D}$ are separated by $\Delta_\mathrm{SD} = 20\lambda$.
Defining $R_\mathrm{S} = |\vec r_0-\vec r_\mathrm{S}|$, $R_\mathrm{D} = |\vec r_0-\vec r_\mathrm{D}|$, and the excess path length $\delta L = R_\mathrm{S}+R_\mathrm{D}-\Delta_\mathrm{SD}$, the product of three Green's functions reduces to
\be
G(\vec r_0,\vec r_\mathrm{S})\,G^{*}(\vec r_\mathrm{D},\vec r_\mathrm{S})\,G(\vec r_0,\vec r_\mathrm{D}) \;\propto\; \frac{e^{i(\omega/c)\,\delta L}}{\Delta_\mathrm{SD}\,R_\mathrm{S}\,R_\mathrm{D}},
\ee
where the phase factor $e^{-i(\omega/c)\Delta_\mathrm{SD}}$ contributed by $G^{*}(\vec r_\mathrm{D},\vec r_\mathrm{S})$ combines with the path factors $e^{i(\omega/c)(R_\mathrm{S}+R_\mathrm{D})}$ to leave the excess-path factor $e^{i(\omega/c)\,\delta L}$, and its amplitude contributes the geometric factor $1/\Delta_\mathrm{SD}$.
The $\mathrm{Re}$ and $\mathrm{Im}$ projections of the remaining factor $e^{i(\omega/c)\delta L}/(R_\mathrm{S} R_\mathrm{D})$ give the monochromatic even and odd kernels,
\begin{align}
K_\mathrm{R}(\vec r_0) \;&\propto\; \frac{\cos\!\big(\omega\,\delta L/c\big)}{R_\mathrm{S}\,R_\mathrm{D}} \;\propto\; \mathrm{Re}[\mathcal{P}], \label{eq:K_R}\\[2pt]
K_\mathrm{I}(\vec r_0) \;&\propto\; \frac{\sin\!\big(\omega\,\delta L/c\big)}{R_\mathrm{S}\,R_\mathrm{D}} \;\propto\; \mathrm{Im}[\mathcal{P}]. \label{eq:K_I}
\end{align}
$K_\mathrm{R}$ is maximal at $\delta L = 0$, the even class, and $K_\mathrm{I}$ vanishes at $\delta L = 0$, the odd class.
Being the two quadratures of a single complex exponential, the cosine and the sine projections exhaust its degrees of freedom, so any kernel in this homogeneous system can only be their linear combination. This is the ballistic expression of the two-class structure.
The area-specific prefactors are powers of $\omega$ that carry no $\delta L$ dependence, see Fig.~1.

For the broadband panels Fig.~2(c, d), with $\omega^2|j_\mathrm{s}(\omega)|^2$ given by a Gaussian spectral envelope $\exp[-(\omega-\omega_0)^2/(2\sigma_\omega^2)]$ of standard deviation $\sigma_\omega$, with $\sigma_\omega/\omega_0 = 0.5$, chosen for concreteness, Eq.~(6) yields
\begin{align}
K_\mathrm{R}^\mathrm{BB}(\vec r_0) \;&\propto\; \frac{\cos\!\big(\omega_0\,\delta L/c\big)\,\exp\!\big[-(\sigma_\omega\,\delta L/c)^2/2\big]}{R_\mathrm{S}\,R_\mathrm{D}}, \label{eq:K_R_BB}\\[2pt]
K_\mathrm{I}^\mathrm{BB}(\vec r_0) \;&\propto\; \frac{\sin\!\big(\omega_0\,\delta L/c\big)\,\exp\!\big[-(\sigma_\omega\,\delta L/c)^2/2\big]}{R_\mathrm{S}\,R_\mathrm{D}}. \label{eq:K_I_BB}
\end{align}
The Gaussian envelope confines sensitivity to $\delta L \lesssim c/\sigma_\omega$, suppressing higher Fresnel zones while preserving the on-ray maximum of $K_\mathrm{R}^\mathrm{BB}$ and the on-ray zero of $K_\mathrm{I}^\mathrm{BB}$.
The dashed curve on each panel of Fig.~2 marks the contour of excess path length $\delta L = \lambda/4$, the first zero of the CW even kernel Eq.~(10).
The on-ray null of the odd kernel is a structural feature of the sine projection, independent of spectral weighting.
Both on-ray descriptors hold in 3D.
In 2D the far-field Green's function carries a phase lag of $\pi/4$ per propagation leg, so the on-ray phase of the kernel oscillation has magnitude $\pi/4$ rather than zero.
The odd kernel then no longer vanishes on the geometric ray, its on-ray value being $\sin(\pi/4)\approx0.71$ of the local oscillation amplitude, and the even kernel is correspondingly reduced from its on-ray maximum~\cite{2002_Spetzler_Trampert_Snieder,2004_Zhou_Dahlen_Nolet}.
Practical 2D inversions of point-source data remove this mismatch at the data level, where the standard line-source transformation applies a half integration together with the compensating $\pi/4$ phase advance~\cite{2014_Forbriger_line_source,2013_Auer_2D_FWI}.
The Re/Im class split itself is dimension-independent.
All four panels in Fig.~2 are evaluated analytically on a $1100\times700$ grid over a 2D cross-section of the source-detector plane.

Unlike the idealized case considered above, practical applications of CWT are better described by a semi-infinite geometry. Although the presence of the boundary reshapes the spatial envelope of the kernel via image sources or reflecting boundary conditions~\cite{2000_Dahlen_banana_doughnut}, the even/odd class split is set by the $\mathrm{Re}/\mathrm{Im}$ quadratures of $\mathcal{P}$ and therefore remains unaffected. In this geometry, a reflector plane or a depth gradient in wave speed extends the reach of the kernel to depth while still preserving the $\mathrm{Re}/\mathrm{Im}$ split, producing the reflected-arrival `rabbit-ear' kernels~\cite{2012_Xu_RWI,2020_Yao_RWI_review,2002_Dahlen_Baig_amplitude} and the curved diving-wave `banana' shapes of finite-frequency tomography~\cite{2000_Dahlen_banana_doughnut,2004_Spetzler_Snieder_Fresnel}, see Supplementary Note~S8 and Supplementary Fig.~S2.

\subsection*{Phase diagram}

Figure~3 compiles published studies from which both transport parameters could be extracted.
Each data point in the diagram is annotated with the first author's name and the year of publication, and is found in the bibliography of this manuscript.
Table~I organizes these data in a table listing specific values for both transport parameters $k\ell_s$ and $L/\ell^*$ and includes a short comment on how they were obtained.
$L$ represents the path spanned by the inversion, such as the borehole separation or two-way surface path in GPR, the imaging depth or array diameter in ultrasound, the vertical canopy extent in TomoSAR, the maximum offset in seismic FWI, the source-receiver offset on volcanic edifices, the source-station distance for the crust, or the source-detector separation in DOT.
Scattering and transport mean free paths $\ell_s$ and $\ell^*$ are used whenever directly quoted in a study.
Below we provide information on the post-processing steps in three cases, marked with $^{\dagger}$, $^{\ddagger}$, and $^{\#}$, where we were unable to extract the parameters directly.

$^{\dagger}$~The crosshole GPR studies listed in Table~I do not provide the scattering mean free path.
We obtain estimates of $k\ell_s$ from the one crosshole GPR survey that separated the scattering quality factor of the direct wave, $Q_\mathrm{sc}=80$--$500$ at $100$~MHz in a dolomite breccia with clasts of $0.3$--$0.6$~m, cf.\ Fig.~10 of Ref.~\cite{2012_Harbi_McMechan_scattering_Q_GPR}.
$k\ell_s=Q_\mathrm{sc}$, and we use it as the analog for the sand and gravel aquifers at the same frequency.
Glacial till and diamict scatter more strongly, $k\ell_s\leq11$--$73$ at the wavenumber of the gravel sites~\cite{2014_Arcone_glacial_till_GPR_diffractions}, which marks the strong-scattering limit of this range.
The transport depth follows from $L/\ell^*\leq L/\ell_s$ with $\ell_s=Q_\mathrm{sc}/k$ at the wavenumber in the corresponding setup.

$^{\ddagger}$~The attenuation of soft tissue is estimated at $\eta$$\approx0.5$~dB/cm/MHz~\cite{2006_Insana_ultrasound_encyclopedia}, while scattering should account for at most $10\%$ of it~\cite{1988_Lyons_Parker_absorption_attenuation}.
From these data we estimate $\ell_s\geq4.34/(0.1\,$$\eta$$f)$ and $k\ell_s=2\pi f\ell_s/c\geq3.5{\times}10^3$ at $c=1540$~m/s, independent of frequency.
The transport depth $L/\ell^*$ is obtained using $\ell_s\leq\ell^*$.

$^{\#}$~Ref.~\cite{2011_Aubry_Derode_soft_tissue} reports the multiple-scattering rate as a function of time, its Fig.~8(c), which reaches $50\%$ at $T\approx46$~\textmu s, i.e., at the depth $z=cT/2=34.5$~mm for $c=1500$~m/s.
$\ell_s$ is obtained from the single-scattering ratio by setting $\exp(-3z/2\ell_s)$ of Eq.~(5) in Ref.~\cite{2024_Goicoechea_ells} to $1/2$, which gives $\ell_s\approx1.5\,z/\ln2=75$~mm.

\begin{table*}[!htbp]
\caption{\label{tab:points}%
Studies plotted in Fig.~3, with the scattering strength $k\ell_s$, the transport depth $L/\ell^*$, and the basis of each value. Values preceded by $\leq$ or $\geq$ are bounds, drawn with arrows in the figure. The symbols $\dagger$, $\ddagger$ and $\#$ refer to the notes in the text.
$\eta$ is the scattering attenuation, $\kappa_e$ the one-way extinction coefficient, $Q_\mathrm{sc}$ the scattering quality factor, $g_0$ the total scattering coefficient, and $v_\mathrm{S}$ the shear-wave speed.}
\scriptsize
\renewcommand{\arraystretch}{0.95}
\setlength{\tabcolsep}{3pt}
\begin{tabular}{p{1.35in}p{1.55in}>{\centering\arraybackslash}p{0.95in}>{\centering\arraybackslash}p{0.75in}p{2.0in}}
\hline\hline
Study & Medium, band & $k\ell_s$ & $L/\ell^*$ & Basis \\
\hline
\textit{GPR} & & & $L/\ell^*\leq L/\ell_s$ & \\
Grimm 2006~\cite{2006_Grimm_GPR_Bishop} & welded tuff, 12.5--50~MHz & $3.9$--$27$ & $\leq1.2$--$9.8$ & $\ell_s=4.34/\eta$, $\eta$ is the scattering attenuation in dB/m from the radar coda envelope, $L=20$--$44$~m \\
Grimm 2007~\cite{2007_Grimm_SEG_Bishop_Tuff_attenuation} & welded tuff, $12$--$180$~MHz effective & $8.8$--$39$ & $\leq3.3$--$5.3$ & Same site, $\eta$ per effective frequency, $L=14$--$44$~m \\
Looms 2010~\cite{2010_Looms_crosshole_GPR_sand} & unsaturated sand, crosshole, $100$~MHz & $(0.8$--$5){\times}10^2$ & $\leq0.05$--$0.28$ & $k\ell_s$ from Ref.~\cite{2012_Harbi_McMechan_scattering_Q_GPR}$^{\dagger}$, $L=5$~m \\
Gueting 2017~\cite{2017_Gueting_Krauthausen_GPR_FWI} & saturated sandy gravel, crosshole, $100$~MHz & $(0.8$--$5){\times}10^2$ & $\leq0.05$--$1.3$ & $k\ell_s$ from Ref.~\cite{2012_Harbi_McMechan_scattering_Q_GPR}$^{\dagger}$, $L=3$--$12$~m \\
Klotzsche 2010~\cite{2010_Klotzsche_GPR_crosshole} & saturated gravel, crosshole, $100$~MHz & $(0.8$--$5){\times}10^2$ & $\leq0.08$--$0.73$ & $k\ell_s$ from Ref.~\cite{2012_Harbi_McMechan_scattering_Q_GPR}$^{\dagger}$, $L=5$--$7$~m \\
Harbi 2012~\cite{2012_Harbi_McMechan_scattering_Q_GPR} & dolomite breccia, crosshole, $100$~MHz & $(0.8$--$5){\times}10^2$ & $\leq0.04$--$1.2$ & $Q_\mathrm{sc}=80$--$500$, $k\ell_s=Q_\mathrm{sc}$, $L=$$8$--$16$~m \\
\hline
\textit{TomoSAR} & & $k\ell_s\geq k/\kappa_e$ & $L/\ell^*\leq\kappa_e L$ & \\
Praks 2012~\cite{2012_Praks_extinction_Lband} & boreal forest, L-band & $\geq3.3{\times}10^2$--$1.5{\times}10^3$ & $\leq0.32$--$2.0$ & $\kappa_e=0.08$--$0.36$~dB/m, $L=17$--$24$~m \\
Papathanassiou 2001~\cite{2001_Papathanassiou_Cloude_RVoG} & temperate forest, L-band & $\geq5.7{\times}10^2$ & $\leq0.92$--$1.4$ & $\kappa_e=0.2$~dB/m, $L=20$--$30$~m \\
Kugler 2015~\cite{2015_Kugler_polinsar_extinction} & boreal, L-band & $\geq(1.1$--$2.3){\times}10^3$ & $\leq0.06$--$0.68$ & $\kappa_e=0.05$--$0.1$~dB/m, $L=5$--$30$~m \\
Kugler 2015~\cite{2015_Kugler_polinsar_extinction} & temperate, L-band & $\geq5.7{\times}10^2$--$1.1{\times}10^3$ & $\leq0.12$--$1.6$ & $\kappa_e=0.1$--$0.2$~dB/m, $L=5$--$35$~m \\
Kugler 2015~\cite{2015_Kugler_polinsar_extinction} & tropical, P-band & $\geq(2.0$--$4.0){\times}10^2$ & $\leq0.70$--$2.1$ & $\kappa_e=0.1$--$0.2$~dB/m, $L=30$--$45$~m \\
\hline
\textit{Seismic} & & $k\ell_s\leq k\ell^*$ & & \\
Eulenfeld 2016~\cite{2016_Eulenfeld_Wegler_basins} & Molasse basin, $3$--$7$~Hz & $\leq(6.5$--$9.9){\times}10^2$ & $0.08$--$0.30$ & $\ell^*=54$--$93$~km, $L=7$--$16$~km, the FWI offsets of Refs.~\cite{2013_Warner_3D_FWI,2009_Virieux_FWI_review} \\
Eulenfeld 2016~\cite{2016_Eulenfeld_Wegler_basins} & Upper Rhine Graben, $3$--$12$~Hz & $\leq17$--$3.0{\times}10^2$ & $0.7$--$7.3$ & $\ell^*=2.2$--$9.7$~km, same $L$ \\
Wegler 2001~\cite{2001_Wegler_Luhr_volcano} & Merapi, $4$--$20$~Hz & $\leq1.3$--$11$ & $10$--$45$ & $Q_\mathrm{sc}=1.3$--$11$, $k\ell^*=Q_\mathrm{sc}$, $L=1$--$3.75$~km \\
Wegler 2003~\cite{2003_Wegler_Vesuvius} & Vesuvius, $2$--$16$~Hz & $\leq1.7$--$13$ & $5$--$20$ & $\ell^*=0.2$~km, $L=1$--$4$~km \\
Yamamoto 2010~\cite{2010_Yamamoto_Sato_Asama} & Asama, $8$--$16$~Hz & $\leq39$--$78$ & $0.4$--$3.3$ & $\ell^*=1.21$~km, $L=0.5$--$4$~km \\
Ib\'a\~nez 2020~\cite{2020_Ibanez_Etna_scattering} & Etna, $4$--$24$~Hz & $\leq10$--$63$ & $3.5$--$25$ & $Q_\mathrm{sc}=10$--$63$, $k\ell^*=Q_\mathrm{sc}$, $L=3.8$--$26$~km \\
Carcol\'e 2010~\cite{2010_Carcole_Sato_MLTWA_Japan} & crust, Japan, $1$--$32$~Hz & $\leq1.9{\times}10^2$--$5{\times}10^3$ & $\leq0.59$--$1.4$ & $\ell^*=70$--$170$~km, $L\leq100$~km \\
Hoshiba 1993~\cite{1993_Hoshiba_MLTWA_Japan} & crust, Japan, $1$--$8$~Hz & $\leq69$--$5.3{\times}10^3$ & $\leq0.23$--$4.7$ & $Q_\mathrm{sc}$ per station as tabulated in Ref.~\cite{2010_Carcole_Sato_MLTWA_Japan}, $k\ell^*=Q_\mathrm{sc}$, $L\leq120$~km \\
Eulenfeld 2017~\cite{2017_Eulenfeld_Wegler_USarray} & crust, contiguous US, $1.5$--$17$~Hz & $\leq1.7{\times}10^2$--$2.0{\times}10^3$ & $\leq0.66$--$1.4$ & $\ell^*=77$--$167$~km~\cite{2019_Sato_GJI_scattering_compilation}, $L\leq110$~km \\
van Laaten 2022~\cite{2022_vanLaaten_MLTWA_Qopen} & crust, $3$--$34$~Hz & $\leq(3.6$--$7.7){\times}10^3$ & $\leq0.11$--$0.77$ & $Q_\mathrm{sc}=(3.6$--$7.7){\times}10^3$, $k\ell^*=Q_\mathrm{sc}$, $L\leq90$~km \\
Gaebler 2015~\cite{2015_Gaebler_Vogtland_scattering} & crust, Vogtland, $3$--$24$~Hz & $\leq5.4{\times}10^2$--$2.4{\times}10^3$ & $0.09$--$0.85$ & $Q_\mathrm{sc}=5.4{\times}10^2$--$2.4{\times}10^3$, $k\ell^*=Q_\mathrm{sc}$, $L=10$--$40$~km \\
Przybilla 2009~\cite{2009_Przybilla_Korn} & crust, southern Norway, $2$--$10$~Hz & $\leq6.5{\times}10^2$--$1.5{\times}10^3$ & $0.25$--$1.0$ & $\ell^*=100$--$200$~km, $L=50$--$100$~km, assumes $v_\mathrm{S}\approx3.5$~km/s~\cite{1995_Christensen_Mooney_crust,1991_Kennett_Engdahl_iasp91} \\
Lacombe 2003~\cite{2003_Lacombe_Lg_coda_France} & crust, France, $2$--$5$~Hz & $\leq(1.1$--$1.3){\times}10^3$ & $0.46$--$5.7$ & $\ell^*=150$--$300$~km, $L=139$--$860$~km \\
Lee 2010~\cite{2010_Lee_Yun_Do_Gyeongsang_basin_MLTWA} & crust, Korea, $1$--$16$~Hz & $\leq3.7{\times}10^2$--$7.6{\times}10^3$ & $0.02$--$0.29$ & $\ell^*=1/g_0$, $g_0=(3.8$--$4.8){\times}10^{-3}$~km$^{-1}$, $L=6$--$60$~km, assumes $v_\mathrm{S}\approx3.5$~km/s~\cite{1995_Christensen_Mooney_crust,1991_Kennett_Engdahl_iasp91} \\
\hline
\textit{Medical ultrasound} & & & $L/\ell^*\leq L/\ell_s$ & \\
Go\"icoechea 2024~\cite{2024_Goicoechea_ells} & liver in vivo, $7.5$~MHz & $(1.2$--$1.5){\times}10^3$ & $\leq0.75$--$1.5$ & $\ell_s=44\pm4$~mm, $L=36$--$59$~mm \\
Aubry 2011~\cite{2011_Aubry_Derode_soft_tissue} & breast in vivo, $4.3$~MHz & $1.3{\times}10^3$ & $\leq0.64$ & $\ell_s=75$~mm from the single-scattering ratio$^{\#}$, $L=48$~mm \\
Wiskin 2020~\cite{2019_Wiskin_breast_ultrasound} & breast and limb FWI, $0.35$--$1.3$~MHz & $\geq3.5{\times}10^3$ & $\leq0.09$--$0.32$ & Attenuation bound on $\ell_s$$^{\ddagger}$, $L=214$~mm \\
Guasch 2020~\cite{2020_Guasch_FWI_brain} & transcranial FWI, $0.1$--$0.85$~MHz & $\geq3.5{\times}10^3$ & $\leq0.02$--$0.21$ & Attenuation bound on $\ell_s$$^{\ddagger}$, $L=210$~mm \\
Sandhu 2015~\cite{2015_Sandhu_ultrasound_FWI} & breast waveform tomography, $0.4$--$1$~MHz & $\geq3.5{\times}10^3$ & $\leq0.09$--$0.23$ & Attenuation bound on $\ell_s$$^{\ddagger}$, $L=200$~mm \\
Lucka 2022~\cite{2022_Lucka_US_FWI} & breast FWI, numerical, to $1.5$~MHz & $\geq3.5{\times}10^3$ & $\leq0.36$ & Attenuation bound on $\ell_s$$^{\ddagger}$, $L=209$~mm \\
\hline
\multicolumn{5}{l}{\textit{DOT and WFS-DOT}} \\
Eggebrecht 2014~\cite{2014_Eggebrecht_HDDOT} & adult head, $750$ and $850$~nm & $9.9{\times}10^2$--$1.6{\times}10^3$ & $8$--$56$ & $\ell^*$ from Ref.~\cite{2012_Eggebrecht_HDDOT_head_model}, $g\approx0.9$, $L=13$--$47$~mm \\
Markow 2025~\cite{2025_Markow_Culver_HDDOT} & adult head, $685$ and $830$~nm & $(8.8$--$9.9){\times}10^2$ & $7$--$58$ & $\ell^*$ from Ref.~\cite{2013_Jacques_tissue_optics}, $g\approx0.9$, $L=6.5$--$40$~mm \\
Vidal-Rosas 2021~\cite{2021_VidalRosas_LUMO_HDDOT} & adult head, $735$ and $850$~nm & $9.9{\times}10^2$--$1.6{\times}10^3$ & $6$--$48$ & As Eggebrecht 2014, $L=10$--$40$~mm \\
Bender 2022~\cite{2022_Bender_coherent_remission} & 2D silicon slab, $1.55$~\textmu m & $74$ & $10$--$25$ & $\ell^*=\ell_s=6.4$~\textmu m, $L=10$--$25\,\ell^*$ \\
\hline
\hline
\end{tabular}
\end{table*}

The diagonal contours of constant $L/\lambda$ in Fig.~3 use $\ell^* \approx 2\pi\ell_s$, so that $L/\lambda = (k\ell_s)(L/\ell^*)$, and they are lines of slope $-1$ in the log-log plane.
The factor $2\pi$, corresponding to the scattering anisotropy of $g\approx 0.84$, is chosen for convenience.
The ratio $\ell^*/\ell_s$ is resolved only in the DOT entries, ${\sim}10$ at $g\approx0.9$~\cite{2013_Jacques_tissue_optics,2010_Durduran_diffuse_optics}, and in the WFS-DOT entry, where it is $1$.
Elsewhere Table~I estimates it as the $\ell_s\leq\ell^*$ bound.
The ovals enclose the points of each group and are shown as a guide for the eye only.

\subsection*{Broadband compression simulations}
\begin{table*}[!htbp]
\caption{The four systems of the OSCAR-BB simulations.
$k\ell_s$ and $\gamma$ are evaluated at the band center. $g$ is the scattering anisotropy, measured on shell in 2D and exact for the top-hat generator in 3D.
The 3D comb has $25$ frequencies $\omega_0(1+\delta)$ with $\delta=0$, $\pm0.00125$, $\pm0.0025$, $\pm0.005$, $\pm1/150$, $\pm0.01$, $\pm1/75$, $\pm0.02$, $\pm2/75$, $\pm0.04$, $\pm4/75$, $\pm5/75$ and $\pm0.08$.
Figure~5 and Supplementary Figs.~S4, S5, S6, and~S7 use its equally spaced subsets centered on $\omega_0$, from $3$ (Fig.~5 and Supplementary Fig.~S4) or $5$ (Supplementary Fig.~S7) frequencies at spacing $0.00125$ to $13$ frequencies at spacing $1/75$.}
\label{tab:SI_systems}
\centering
\footnotesize\setlength{\tabcolsep}{2pt}
\begin{tabular}{l>{\centering\arraybackslash}p{2.7cm}>{\centering\arraybackslash}p{2.7cm}>{\centering\arraybackslash}p{3.2cm}>{\centering\arraybackslash}p{3.2cm}}
\hline\hline
 & 2D, $k\ell_s\approx80$ & 2D, $k\ell_s\approx150$ & 3D, $k\ell_s\approx300$ & 3D, $k\ell_s\approx600$ \\
\hline
Disorder & Holes, $f_0=0.1$ & Holes, $f_0=0.05$ & Top-hat, $q_c=14$~\textmu m$^{-1}$ & Top-hat, $q_c=14$~\textmu m$^{-1}$ \\
Vacuum wavelength, index & $1.55$~\textmu m, $2.85$ & $1.55$~\textmu m, $2.85$ & $0.85$~\textmu m, $1.35$ & $0.85$~\textmu m, $1.35$ \\
Slab depth $L$, width & $150$, $99.2$~\textmu m & $150$, $99.2$~\textmu m & $189$, $54$~\textmu m & $378$, $54$~\textmu m \\
$\ell_{s0}=\ell_s(\omega_0)$, Supplementary Fig.~S3 & $7.4$~\textmu m & $13$~\textmu m & $30$~\textmu m & $60$~\textmu m \\
Transport depth $\gamma=L(1-g)/\ell_{s0}$ & $20$ & $11$ & $3.1$ & $3.1$ \\
$g$ & $0.03$ & $0.06$ & $0.51$ & $0.51$ \\
Comb frequencies, spacing, span $B$ & $21$, $0.01$, $0.2$ & $21$, $0.01$, $0.2$ & $25$, see caption, $0.16$ & $25$, see caption, $0.16$ \\
Mask widths $\alpha$ & $4$, $12$ & $4$, $12$ & $4$, $12$ & $4$, $12$ \\
Realizations $N_\mathrm{rlz}$ & $10$ & $10$ & $10$ & $10$ \\
\hline\hline
\end{tabular}
\end{table*}

All simulations of Fig.~5 and of Supplementary Note~S10 use the four systems listed in Table~II.
The disorder in 2D media with background of index $2.85$ is introduced by a random placement of air holes of radii $40$--$65$~nm at a minimum separation of $50$~nm with the nominal filling fraction $f_0$.
The monochromatic solutions for the out-of-plane electric field, the TE configuration of Sec.~II\,C, are obtained with the frequency-domain solver MESTI~\cite{2022_Lin_MESTI} at $20$ grid points per vacuum wavelength, cf.~Ref.~\cite{2026_Jara_OSCAR_compression}.
The 3D media represent a continuous refractive-index disorder $n_0+\delta n(\vec r)$ with a spectral top-hat generator of cutoff $q_c$ and amplitude $\delta n=0.045$ and $0.0318$ for the $k\ell_s\approx300$ and $600$ systems, solved for the vector electromagnetic field with the modified Born series solver~\cite{2016_Osnabrugge_bornseries} at three grid points per medium wavelength, cf.~Ref.~\cite{2026_Jara_OSCAR_compression}.
In both dimensions the slab is illuminated by a plane wave at $z=0$ with periodic transverse boundaries, and in 3D the analysis uses the $x$ component of the field, and the adjoint field of each frequency is launched from $z=L$ by conjugating the outgoing wavefront.
The mask of each system is common to all realizations of the disorder and is set by the shell radius $k(\omega)$ and width $1/\ell_s(\omega)$ at every frequency of the comb.
The radius follows from the dispersion relation, $k(\omega)=n_\mathrm{eff}\,\omega/c_0$, with one effective index per medium, the background index in 3D and the value measured at $\omega_0$ from the ensemble-averaged spectrum in 2D.
To account for grid dispersion in 2D, the mask is applied in terms of the finite-difference wavenumber of the five-point stencil, in which the shell is a circle~\cite{2026_Jara_OSCAR_compression}.
The width is $\alpha/\ell_s(\omega)$, where $\ell_s(\omega)$ is the fit over the harmonic means of Supplementary Fig.~S3.
A pulse of normalized bandwidth $B$ is a comb of span $B$ with a Gaussian spectral envelope of full width at half maximum $B/2$, so that the edge frequencies carry $6\%$ of the peak amplitude.
The fields are stored in one of two protocols.
In the per-frequency protocol the field at each frequency is compressed on its own shell.
The pulse is a sum over the comb $\{\omega_j\}$, $E_0(\vec r,t)=\mathrm{Re}\sum_j s_j\,E_0(\vec r,\omega_j)\,e^{-i\omega_j t}$, with the source weights $s_j$ of its frequencies, and the stored field at $\omega_j$ and its reconstruction are
\begin{align}\label{eq:SI_protocol_F}
E_0(\vec r,\omega_j)\;&\to\;\widetilde E_c^{\mathrm F}(\vec k,\omega_j)=P_j\,\widetilde E_0(\vec k,\omega_j)\,\frac{\|\widetilde E_0(\omega_j)\|}{\|P_j\widetilde E_0(\omega_j)\|}\nonumber\\
&\to\;E_c^{\mathrm F}(\vec r,t)=\mathrm{Re}\sum_j s_j\,E_c^{\mathrm F}(\vec r,\omega_j)\,e^{-i\omega_j t},
\end{align}
where $\widetilde E$ is the spatial Fourier transform, $P_j$ the mask of width $\alpha$ on the shell of $\omega_j$, and the ratio of the norms, $\|\widetilde E\|^2=\int|\widetilde E|^2\,d\vec k$, the power-restoring factor of Ref.~\cite{2026_Jara_OSCAR_compression} at frequency $\omega_j$.
The labels F and S distinguish the per-frequency and the per-snapshot protocols.
In the per-snapshot protocol the fields at a given time are compressed as they are computed,
\be\label{eq:SI_protocol_S}
E_0(\vec r,t)\;\to\;\widetilde E_c^{\mathrm S}(\vec k,t)=P_{\mathrm S}\,\widetilde E_0(\vec k,t)\,\frac{\|\widetilde E_0(t)\|}{\|P_{\mathrm S}\widetilde E_0(t)\|},
\ee
with one power-restoring factor per snapshot and the same mask $P_\mathrm S$ for every snapshot.
The mask $P_\mathrm S$ is the annulus that spans the shells of all frequencies of the band, Eq.~(S56) of Supplementary Note~S10.
To quantify the fidelity of reconstruction of a compressed pulse, we generalize the CW overlap error of Ref.~\cite{2026_Jara_OSCAR_compression} to time-dependent fields
\be\label{eq:SI_overlap_error}
\epsilon_\phi=1-\frac{\left|\langle E_c,E_0\rangle\right|}{\|E_c\|\,\|E_0\|},\qquad
\langle A,B\rangle=\int_0^T\!dt\int d\vec r\,A\,B,
\ee
where $E_0(\vec r,t)$ is the original real time-domain field and $E_c(\vec r,t)$ the field reconstructed from the compressed data.
$T$ is the (alias) period of the comb, $\|A\|^2=\langle A,A\rangle$, and the spatial integral runs over the stored volume used for reconstruction, the slab $0<z<L$ in both dimensions.
The per-snapshot protocol is evaluated at $64$ or $128$ equally spaced times within one period $T=2\pi/\Delta\omega_\mathrm{step}$ of the comb, where $\Delta\omega_\mathrm{step}$ is its frequency spacing.
In both cases no two frequencies of the comb alias onto each other on the time grid, so the sum over the period replaces the time integral exactly.
The kernel errors are evaluated over the slab after a boundary trim of $3/\Delta k=3\ell_s/\alpha$ at each face, the convention of Ref.~\cite{2026_Jara_OSCAR_compression}.
There, the Gibbs layer of the sharp mask was measured to lie within $1/\Delta k$ of a face.
Every quantity is the mean over $N_\mathrm{rlz}=10$ realizations of the disorder, with the standard deviation over realizations as the error bar where shown.
In the 3D $k\ell_s\approx600$ system the forward and the adjoint field of each realization enter Fig.~5 as separate samples, so $N_\mathrm{rlz}=10$ stands for $20$ samples there.

\vspace{0.35cm}
\section*{Acknowledgements}
The author thanks Pablo Jara for discussions on the OSCAR compression algorithm, and Arthur Goetschy and Hui Cao for helpful discussions.
The simulations were performed on the Mill high-performance computing cluster of Missouri University of Science and Technology~\cite{2024_Mill_HPC_cluster}. A part of the computation for this work was performed on the high-performance computing infrastructure operated by Research Support Solutions in the Division of IT at the University of Missouri, Columbia, MO~\cite{2025_Hellbender_HPC_cluster}.

\section*{Competing interests}
A.Y. and P.~Jara are named inventors on a patent application related to the OSCAR algorithm described in this paper, filed by Missouri University of Science and Technology.

\bibliographystyle{naturemagdoi}
\bibliography{Sensitivity_analogies}
\end{document}